\documentclass[11pt]{article}

\usepackage[margin=1in]{geometry}
\usepackage{amsmath,amssymb,bm}
\usepackage{mathtools}
\usepackage{graphicx}
\graphicspath{{figs/}}
\usepackage{natbib}
\setcitestyle{authoryear,square,citesep={;}}
\let\cite\citep            
\usepackage[colorlinks=true,allcolors=blue]{hyperref}
\providecommand{\keywords}[1]{\par\medskip\noindent\textbf{Keywords:}\ #1\par}
\usepackage[section]{placeins}   

\providecommand{\Tr}{\mathrm{Tr}}
\providecommand{\ket}[1]{|#1\rangle}
\providecommand{\bra}[1]{\langle #1|}
\providecommand{\expect}[1]{\langle #1 \rangle}
\providecommand{\norm}[1]{\|#1\|}

\begin{document}

\title{Preparation-Space Diagnostics and Logical Information Loss\\
in a Driven Kerr-Cat Qubit}

\author{S. Wiggins\\[4pt]
\normalsize Hetao Institute of Mathematics and Interdisciplinary Sciences, Shenzhen, China\\
\normalsize School of Mathematics, University of Bristol, Bristol BS8 1UG, United Kingdom}

\date{}

\maketitle

\begin{abstract}
A Kerr-cat qubit encodes a logical bit in the two wells of a
parametrically driven nonlinear oscillator, and a logic gate is a
transient change of the drive.  In the phase plane the gate deforms the
double well and can split its separatrix into a turnstile that carries
trajectories across the dividing surface between the wells; the same
pulse, acting on the quantum oscillator, can corrupt the encoded bit.  We
study this process over a disk of coherent-state preparations, comparing
classical phase-space transport diagnostics with the open-system quantum
outcome on a common domain so that the two can be compared point by point.
The central finding is that the corruption depends on the full temporal
protocol, not on pulse strength alone: a sudden quench erases the bit,
whereas a smooth ramp of the same peak amplitude largely preserves it.  A
finite-time sensitivity field locates the classical transport boundary,
and a Loschmidt echo evaluated near the end of the gate predicts the much
later quantum outcome.  Sweeps of pulse amplitude and width, of cat size,
and of engineered two-photon dissipation map where the classical transport
picture predicts the quantum loss of the bit and where it does not.
\end{abstract}


\section{Introduction}
\label{sec:intro}

A Kerr-cat qubit is built from a single nonlinear oscillator (in practice a superconducting microwave circuit) whose Hamiltonian contains a Kerr
term, an energy quadratic in the photon number, together with a two-photon
parametric pump that adds and removes photons in
pairs~\cite{Grimm2020,Goto2016,Puri2017}.  The pump is called
\emph{parametric} because the external tone modulates a parameter of the
circuit at twice the oscillator frequency rather than forcing the
oscillator directly; we use ``pump'' and ``parametric drive''
interchangeably.  In a frame rotating at half the
pump frequency, the combined effect of these two terms is an effective
double-well structure in the oscillator's phase plane: the two wells are
centered on the coherent states $\ket{\pm\alpha_0}$, minimum-uncertainty
wavepackets of opposite displacement.  The qubit encoding identifies the
two wells with the two logical states.  A \emph{bit-flip error} is a
transition of the state from one well to the other.  In the idling
configuration (constant pump) such transitions are exponentially
suppressed in the well separation~\cite{Cochrane1999,Mirrahimi2014},
which is the principal attraction of the architecture, and bit-flip times
of seconds have been demonstrated~\cite{Berdou2023,Reglade2024}.  Gate
operations, however, require deliberately and transiently changing the
pump, and a pulse that deforms the double well can drive the state across
the barrier.  Identifying which pulses, acting on which initial states,
cause such crossings is a central design question for this hardware.

The simplest description of the process is classical.  The expectation
value of the oscillator's annihilation operator,
$\alpha(t) = \expect{a}_t = x(t) + iy(t)$, obeys (in the mean-field approximation, which discards correlations between fluctuations) a pair
of ordinary differential equations in the phase plane $(x, y)$
(Sec.~\ref{sec:model}).  For a constant pump $p_0$ this two-dimensional
flow realizes the double-well picture concretely: two stable equilibria
(the bottoms of the wells), separated by an equilibrium of saddle type at
the origin.  Because the physical device loses photons at a rate
$\kappa$, the flow is dissipative, and each stable equilibrium attracts a
basin of initial conditions.  The set of initial conditions whose
trajectories approach the saddle (in the language of dynamical systems, the \emph{stable manifold} of the saddle) forms the boundary between the
two basins: initial conditions on one side settle into the right well,
those on the other side into the left.  We refer to the vertical line
$x = 0$ through the saddle as the \emph{dividing surface} between the two
logical regions, and to a classical trajectory that ends at $x(T) < 0$,
having started in the right region, as having \emph{leaked}.

Ref.~\cite{WigKC} analyzed what a gate pulse does to this picture.  When
the pump is pulsed the equations of motion become explicitly time
dependent, and the stable and unstable manifolds of the saddle, which in
the conservative limit coincide along a single separatrix curve, split
apart and can intersect one another transversely.  Melnikov
theory~\cite{Melnikov1963,wiggins2003} provides a computable function
$M(t_0)$ that measures, to first order in the perturbation, the signed
distance between the split manifolds as a function of the reference time
$t_0$ at which the distance is evaluated.  Simple zeros of $M$ correspond
to intersections of the manifolds, and the region enclosed between the
two manifolds and two successive intersection points (a \emph{lobe}) is
carried bodily from one side of the dividing surface to the other.  This
is the turnstile mechanism of phase-space
transport~\cite{RomKedar1990}.  Its practical output is a threshold in
the pulse parameters: for a Gaussian pulse of amplitude $A$ and width
$\sigma$, Melnikov theory predicts transport, to first order, when
$\max_{t_0} M(t_0; A, \sigma) > 0$, and the lobe identifies, to leading
order, which initial conditions cross.

This paper examines the same process from the standpoint of the initial
state rather than the pulse.  We define the \emph{preparation space} as
the set of initial quantum states available to the experimenter.  Here we
take it to be the family of coherent states $\ket{\alpha_0}$: a coherent
state is the minimum-uncertainty Gaussian wavepacket centered at the
phase-plane point $\alpha_0 = x_0 + iy_0$, it is the state produced by
displacing the oscillator vacuum, and it is the closest quantum analogue
of a classical initial condition.  The preparation space is then the
two-dimensional disk of center points $(x_0, y_0)$ surrounding the
right-hand well.

Every preparation-resolved diagnostic in this paper (classical or quantum) is expressed as a \emph{field} on this disk, and the word is meant in its plain
mathematical sense: a rule attaching one number to each preparation.
This is the methodological core of the paper.  A single classical
trajectory and a single quantum evolution are objects of different
kinds and admit no direct comparison; but a classical diagnostic and a
quantum diagnostic that are both fields on the same two-dimensional
space can be compared globally, point by point, and the degree of
agreement summarized by a single number.  For that number we use the
Pearson correlation coefficient $r$~\cite{Pearson1895} throughout: $r = +1$ for two fields
that rise and fall together perfectly, $-1$ for fields in perfect
opposition, and $0$ for unrelated fields.  For two fields $u, v$ sampled over the grid,
$r = \sum_i (u_i - \bar u)(v_i - \bar v) / \sqrt{\sum_i (u_i - \bar u)^2 \sum_i (v_i - \bar v)^2}$.  The question throughout is:
which preparations are vulnerable to a given gate pulse, and which
computable field best identifies them?

On the classical side we consider two such fields.  The first is the
\emph{finite-time sensitivity}.  Let $\Phi^T$ denote the trajectory map of the
classical equations: the function that carries an initial condition
$(x_0, y_0)$ to the solution $(x(T), y(T))$ at time $T$.  Its Jacobian
$D\Phi^T$ is the $2 \times 2$ matrix of partial derivatives
$\partial(x(T), y(T))/\partial(x_0, y_0)$, computed by integrating the
linearized (variational) equations along each trajectory, and the
sensitivity is its spectral norm $\norm{D\Phi^T}_2$: the largest singular
value of the matrix, equal to the maximal factor by which an
infinitesimal displacement of the initial condition is amplified after
time $T$.  The second field is the \emph{Lagrangian descriptor}~(LD),
defined as the arc length of the trajectory over the time
window~\cite{Madrid2009,Mancho2013}.  LDs were introduced as a practical
means of revealing invariant manifolds in time-dependent flows:
manifolds appear as ridges or abrupt changes in the LD field, because
trajectories on opposite sides of a manifold have qualitatively different
histories.  In the companion paper~\cite{WigJCP} both fields were
computed over a preparation space of Gaussian wavepackets in a driven
double-well model; the central finding there was that the sensitivity
field and the gradient of the LD field, although organized by the same
phase-space structures, correlate only weakly with one another ($r \approx 0.10$); they are complementary diagnostics, not
interchangeable ones.  The present paper tests whether, and in what form,
that conclusion carries over to the Kerr-cat model, where the classical
flow is dissipative and the relevant quantum evolution is an open-system
one.

On the quantum side, the outcome must be computed from the full
open-system dynamics, because photon loss is intrinsic to the device and
is part of what stabilizes the wells.  The standard description is the
\emph{Lindblad master equation}~\cite{Lindblad1976,GKS1976}, the evolution equation for the density
operator $\rho$ of a system weakly coupled to an environment; for the
Kerr-cat it consists of the Hamiltonian commutator term plus a dissipator
representing single-photon loss at rate $\kappa$
(Eq.~\eqref{eq:lindblad}).  The quantum outcome field is the
\emph{left-half-plane occupation}
$P_\mathrm{left}(T) = \Tr[\rho(T)\,\Pi_{X<0}]$: the probability that a
measurement of the position-like quadrature
$\hat{X} = \tfrac{1}{2}(a + a^\dagger)$ at the final time finds the
oscillator on the left side of the dividing surface.  $P_\mathrm{left}$
is the direct quantum analogue of the classical leak indicator: it is
near $0$ for a state localized in the right well, near $1$ for a state
localized in the left well, and near $\tfrac{1}{2}$ for a state spread
evenly across both.  The last value is the
signature not of transport but of the scrambling of the information about
which well the oscillator occupies, that is, of the logical bit
itself; we refer to this as the loss of \emph{which-well information}.

To connect the classical and quantum fields we use two further
diagnostics.  The first is the out-of-time-order correlator
(OTOC)~\cite{Larkin1969,Maldacena2016}.  For an initial state
$\ket{\alpha_0}$ we take
$C(T, \alpha_0) =
\bra{\alpha_0}[\hat{X}_H(T), \hat{Y}]^\dagger
[\hat{X}_H(T), \hat{Y}]\ket{\alpha_0}$,
the squared commutator of the Heisenberg-evolved quadrature
$\hat{X}_H(T)$ with the conjugate quadrature
$\hat{Y} = \tfrac{1}{2i}(a - a^\dagger)$.  (The literature writes the
OTOC either as a four-point correlation function with operators out of
time order or, equivalently for our purposes, as this squared
commutator; we compute the squared-commutator form, and ``OTOC'' refers
to it throughout.)  The commutator measures how strongly an operation
performed at time zero affects a measurement made at time $T$; in the semiclassical limit, $C$ reduces to
$\hbar^2\bigl(\partial x(T)/\partial x_0\bigr)^2$, the square of one
entry of the classical Jacobian $D\Phi^T$.  The OTOC is therefore the
natural quantum extension of the classical sensitivity field.  In chaotic systems its early-time growth is exponential at a rate matching
the classical Lyapunov exponent~\cite{ChavezCarlos2019}.  The Kerr oscillator does not, by contrast, exhibit a sustained
exponential OTOC-growth window in the regime we study.  The conservative (undamped)
Kerr oscillator is integrable, and although the pulsed drive splits the
separatrix (the homoclinic tangling that the Melnikov analysis of Sec.~\ref{sec:classical} quantifies) Ref.~\cite{AlmasriReboiro} computed
OTOCs for the closely related non-Hermitian quadratic (Swanson)
Hamiltonian~\cite{Swanson2004} and for Kerr and driven extensions of it,
and found $\lambda_L = 0$, i.e.\ polynomial rather than exponential
commutator growth, throughout the parameter regime relevant here (we work
at resonance, far from the exceptional points where that reference finds
enhanced dephasing).  The absence of a sustained exponential-growth window
has a practical consequence: any correspondence between OTOC and classical
sensitivity must be sought directly, field against field, and (as we find in Appendix~\ref{app:otoc}) it does not survive.  A second consequence concerns
what we will call the \emph{background} evolution: the dynamics
generated by the constant pump and the Kerr term, which run whether or
not a gate is applied.  The OTOC grows (polynomially) under the
background alone, and this growth turns out to set the time window
within which the gate's contribution carries spatial information at
all.  We therefore introduce the \emph{gate-induced} OTOC,
$\Delta C = C_\mathrm{pulsed} - C_\mathrm{static}$: the difference
between the OTOC computed with the pulse and the OTOC computed under
the background alone, which isolates the contribution of the gate.

The second diagnostic is the Loschmidt echo
(LE)~\cite{Peres1984,Jalabert2001}, defined here (in a form adapted to open systems (Sec.~\ref{sec:loop})) as the quantum fidelity
$F\bigl(\rho_\mathrm{pulsed}(T), \rho_\mathrm{static}(T)\bigr)$ between
the two states obtained by evolving the \emph{same} preparation with and
without the pulse.  The OTOC and the LE answer different questions.  The
OTOC is a property of operator dynamics under a single evolution: it
measures how far an initial perturbation spreads.  The LE compares two
evolutions of the state itself: it measures how much the gate changed
the state relative to what free evolution would have done.
$F \approx 1$ means the pulse barely disturbed the preparation; small
$F$ means strong gate-induced disturbance.

Three of these are preparation-resolved fields we rely on: the classical
sensitivity, $P_\mathrm{left}$, and the Loschmidt echo $F$; the
gate-induced OTOC $\Delta C$ is an \emph{exploratory} diagnostic that we
test and ultimately record as a negative result (Appendix~\ref{app:otoc}).
The Lagrangian descriptor (the classical companion of the sensitivity,
Sec.~\ref{sec:classical}) plays a supporting role.  Distinct from all of
these are the \emph{logical-bit} diagnostics (the trace distance between the two logical inputs and its idle-normalized retention $R_D$) which are not fields on the preparation disk but properties of the
encoded bit, and which carry the paper's central result.
The preparation-resolved fields are computed jointly on the same grid, for
the same model and pulse; the logical-bit diagnostics are evaluated for the
two nominal codeword preparations.  Both sets are then re-computed across
sweeps of the pulse amplitude, the pulse width, the single-photon loss
rate, and an added two-photon dissipation channel.  The numerical components check one
another: the mean field of the Lindblad equation reproduces the
classical equations of motion to the expected accuracy
(Sec.~\ref{sec:model}); the zeros of the independently computed Melnikov
function fall at the base of the classical lobe observed in the
simulations (Sec.~\ref{sec:classical}); and the time-ordered
Heisenberg-picture propagation used for the OTOC is validated against
Schr\"odinger-picture state propagation by the agreement of
$\Tr[\hat X_H(t)\,\rho_0]$ with $\Tr[\hat X\,\rho(t)]$ for random
preparations, a check that fixes the operator time-ordering, in
addition to the per-step adjoint identity
$\Tr[A^\dagger\mathcal{L}(\rho)] =
\Tr[(\mathcal{L}^\dagger A)^\dagger\rho]$
(Appendix~\ref{app:numerics}).

The central result is that the gate-induced corruption of the logical bit
depends critically on the full temporal protocol.  A sudden quench to
the pulse amplitude erases the bit: the trace distance between the two
logical inputs collapses to $D_{\mathrm{tr}} = 0.013$, and projection onto
the logical subspace shows a near-even incoherent mixture of the
wells, while a full smooth Gaussian with the same peak amplitude
suppresses the loss,
retaining about half the idle distinguishability ($R_D = 0.46$).  The
classical sensitivity field locates the transport boundary; the quantum
left-occupation tracks classical transport directly ($r = +0.61$ under the
quench); and a Loschmidt echo evaluated near the pulse end predicts the
much later outcome in the quench regime ($r = -0.94$).  Crucially, this left-occupation \emph{field} tracks the classical
partition even though the logical \emph{bit} is erased: spatial
outcome prediction and logical preservation are distinct.  Across the
amplitude and width sweeps the classical-leak maximum coincides with the
logical-distinguishability minimum at the sampled resolution.  The erasure weakens as the oscillator
is made more semiclassical; an engineered pair-loss stabilizer drives the
classical leak fraction to zero while raising the gate-induced retention
from $R_D = 0.02$ to $0.94$, leaving only a finite-time residual outside
the corresponding first-moment mean-field model; and the operator-growth
(OTOC) diagnostic we had hoped would bridge the classical and quantum
pictures does not furnish a robust correspondence, and we record it as a
negative result.

The remainder of the paper is organized as follows.
Section~\ref{sec:model} specifies the classical and quantum models, the
preparation grid, and the outcome observable.
Section~\ref{sec:classical} presents the classical sensitivity field and
its coincidence with the transport boundary.  Section~\ref{sec:quantum}
presents the quantum left-half-plane occupation, the gate-induced
vulnerability field, and robustness to both dissipation channels including
the engineered stabilizer.  Section~\ref{sec:protocol} establishes the
protocol dependence (the quench-versus-smooth contrast measured by the logical trace distance) and the Loschmidt echo as a quench-regime
early-warning diagnostic.  Section~\ref{sec:catsize} scans the cat size.
Section~\ref{sec:regimes} presents the amplitude and width sweeps that
organize the results into a regime map.  Section~\ref{sec:discussion}
collects the correlation structure, relates it to Refs.~\cite{WigJCP}
and~\cite{AlmasriReboiro}, and states the design implications and
limitations.  Section~\ref{sec:conclusion} concludes.  The operator-growth
(OTOC) analysis is collected in Appendix~\ref{app:otoc} as a tested but
unsuccessful bridge.

\section{Model and Setup}
\label{sec:model}

\subsection{Classical equations of motion}

The semiclassical model is taken from Ref.~\cite{WigKC}.  In a frame
rotating at half the pump frequency and at resonance, the mean-field
amplitude $\alpha(t) = x(t) + iy(t)$ obeys
\begin{align}
  \dot{x} &= -\tfrac{\kappa}{2}\,x + \bigl(p(t) + K(x^2 + y^2)\bigr) y,
  \label{eq:cl_x} \\
  \dot{y} &= \bigl(p(t) - K(x^2 + y^2)\bigr) x - \tfrac{\kappa}{2}\,y.
  \label{eq:cl_y}
\end{align}
Here $K$ is the Kerr nonlinearity, $\kappa$ the single-photon loss rate,
and $p(t)$ the parametric pump amplitude.  The terms proportional to
$p(t)$ describe the two-photon drive; the cubic terms come from the Kerr
nonlinearity; the terms proportional to $\kappa/2$ are the mean-field
damping due to photon loss.  The gate is modeled as a Gaussian pulse
superimposed on the constant idling pump,
\begin{equation}
  p(t) = p_0 + A\exp\!\bigl(-(t-t_c)^2 / 2\sigma^2\bigr),
  \label{eq:pulse}
\end{equation}
with amplitude $A$, width $\sigma$, and center time $t_c$.  Throughout we
use the physical parameter values of Ref.~\cite{WigKC} (Table~1 there):
$K = \kappa = 1$, $p_0 = 1.5$; the reference pulse is $A = 6$,
$\sigma = 0.3$, $t_c = 0$; the final
time is $T = 8$ unless stated otherwise.  Times are quoted in units of
$1/K$, the natural timescale set by the Kerr nonlinearity; we refer to
$1/K$ as a \emph{Kerr time}.  One timing convention should be stated
plainly: integration always begins at $t = 0$, so for a pulse centered
at $t_c = 0$ the system experiences only the \emph{decaying half} of
the Gaussian: the protocol is a half-pulse whose effective end lies
near $t \approx 3\sigma$ (hence ``pulse end'' near $t \approx 1$ for
the reference width).  The late-pulse cases ($t_c = 4$) contain the
full Gaussian inside the integration window.

For the constant pump the flow has the double-well structure described
in the Introduction.  There are two stable equilibria, one in each half
plane, lying at $(\pm\sqrt{p_0/K},\,0)$ on the real axis in the
conservative limit; with the strong dissipation used here ($\kappa = K$)
they rotate by about $10^\circ$ to $(x,y) \approx (1.17, 0.20)$ and its
reflection, the standard situation in which the logical code states are
the conservative cat states and loss is a small perturbation.  Between
them lies an equilibrium of saddle type at the origin.  Each stable
equilibrium attracts a basin of initial conditions, and the boundary
between the two basins is the stable manifold of the saddle.  We call
the line $x = 0$ the dividing surface (the perpendicular bisector of the two wells, and the natural bit-flip readout) and we say a trajectory
started in the right region has \emph{leaked} if $x(T) < 0$.  In the
conservative limit $\kappa = 0$ the stable and unstable manifolds of the
saddle coincide along the homoclinic (separatrix) orbit
\begin{equation}
  x_h(t) = \sqrt{2p_0/K}\,\frac{\cosh p_0 t}{\cosh 2p_0 t},
  \qquad
  y_h(t) = \sqrt{2p_0/K}\,\frac{\sinh p_0 t}{\cosh 2p_0 t},
  \label{eq:separatrix}
\end{equation}
which is the unperturbed orbit entering the Melnikov integral used in
Secs.~\ref{sec:classical} and~\ref{sec:regimes}.

\subsection{Quantum model}
\label{sec:qmodel}

The quantum Hamiltonian whose mean field reproduces
Eqs.~\eqref{eq:cl_x}--\eqref{eq:cl_y} is
\begin{equation}
  H(t) = \frac{K}{2}\,a^{\dagger 2}a^2 -
         \frac{p(t)}{2}\bigl(a^{\dagger 2} + a^2\bigr),
  \label{eq:Hquantum}
\end{equation}
where $a$ and $a^\dagger$ are the annihilation and creation operators.
The first term is the Kerr energy: it is diagonal in the photon-number
basis with eigenvalues $\tfrac{K}{2}n(n-1)$, penalizing photon number
quadratically.  The second term is the standard squeezing (two-photon) drive of the
Kerr-cat encoding; with this sign it places the two wells on the real axis
at $\pm\sqrt{p_0/K}$, so that the $x = 0$ readout is their exact
perpendicular bisector.  In competition with the Kerr term it produces the
double-well structure.

Photon loss is included through the Lindblad master equation
\begin{equation}
  \dot{\rho} = -i[H(t), \rho]
  + \kappa\,\mathcal{D}[a]\rho,
  \qquad
  \mathcal{D}[L]\rho \equiv L\rho L^\dagger
  - \tfrac{1}{2}\{L^\dagger L, \rho\},
  \label{eq:lindblad}
\end{equation}
where $\mathcal{D}[a]$ is the standard dissipator describing the loss of
single photons to the environment at rate $\kappa$, and
$\{\cdot,\cdot\}$ is the anticommutator.  Taking the expectation of $a$
in Eq.~\eqref{eq:lindblad} and factorizing third moments
($\expect{a^\dagger a a} \to |\expect{a}|^2 \expect{a}$) recovers
Eqs.~\eqref{eq:cl_x}--\eqref{eq:cl_y}; we verify this numerically by
propagating a coherent state under the constant pump and comparing
$\expect{a}(t)$ with the classical solution, finding agreement up to the
expected quantum correction (the unfactorized part of the third moment).

\paragraph{Two-photon dissipation.}
Physical cat-qubit devices are stabilized not only by single-photon loss
but by \emph{engineered two-photon dissipation}: a coupling to an
auxiliary mode that removes photons from the storage oscillator in
pairs~\cite{Mirrahimi2014,Berdou2023,Reglade2024}.  In
Sec.~\ref{sec:twophoton} we therefore also consider the extended master
equation
\begin{equation}
  \dot{\rho} = -i[H(t), \rho]
  + \kappa\,\mathcal{D}[a]\rho
  + \kappa_2\,\mathcal{D}[a^2]\rho,
  \label{eq:lindblad2}
\end{equation}
in which the dissipator $\mathcal{D}[a^2]$ removes photon pairs at rate
$\kappa_2$.  We emphasize that $\mathcal{D}[a^2]$ models
\emph{phenomenological pair loss}: the reservoir-engineered dissipator
used to stabilize dissipative cat qubits is
$\mathcal{D}[a^2 - \alpha_{\mathrm{cat}}^2]$, which removes pairs
\emph{and} pins the steady states to the wells; we compare the two
explicitly in Sec.~\ref{sec:dpleft}.  Its mean-field contribution is a nonlinear damping term:
taking expectations as above adds $-\kappa_2 |\alpha|^2 \alpha$ to
$\dot{\alpha}$, i.e.\ $-\kappa_2(x^2{+}y^2)\,x$ and
$-\kappa_2(x^2{+}y^2)\,y$ to Eqs.~\eqref{eq:cl_x} and~\eqref{eq:cl_y}
respectively.  This mean-field correspondence is verified numerically in
the same way as above, and the classical comparison runs of
Sec.~\ref{sec:twophoton} use the correspondingly modified vector field.

\subsection{Preparation grid, truncation, and the outcome observable}
\label{sec:prep}

Every preparation-resolved field in this paper is computed on the same
grid of coherent-state initial conditions $\ket{\alpha_0}$, whose centers
$\alpha_0 = x_0 + iy_0$ lie on a Cartesian grid covering the disk of
radius $R = 0.8$ centered at $(\sqrt{p_0/K}, 0) \approx (1.22, 0)$.  The
center is the conservative cat amplitude rather than the dissipative
equilibrium $(1.17, 0.20)$; this convention matches the preparation
ring of Ref.~\cite{WigKC}, and both points lie well inside the disk.
The disk surrounds the right-hand well and reaches toward the dividing
surface; it contains the 150-point preparation ring used in Appendix~B.2
of Ref.~\cite{WigKC}, so the classical results below can be checked
directly against that work.  The joint classical--quantum computations
use a $41 \times 41$ grid, of which the $1256$ points inside the disk are
retained; the purely classical fields of Sec.~\ref{sec:classical}, whose
individual solves are cheap, are refined on a $121 \times 121$ grid
($11287$ interior points).  All quoted statistics are taken over the
interior points of the relevant grid.

Quantum computations represent the oscillator in the photon-number
(Fock) basis truncated at $N = 24$ levels; that is, the state is
expanded in $\ket{0}, \ldots, \ket{23}$ and contributions beyond are
discarded.  The truncation is validated in two ways: the population in
the top three Fock levels remains below $10^{-16}$ in all runs reported,
and repeating the reference computation at $N = 32$ changes no quoted
quantity at the precision given.

The quantum outcome observable is the left-half-plane occupation
\begin{equation}
  P_\mathrm{left}(T) = \Tr\bigl[\rho(T)\,\Pi_{X<0}\bigr],
  \qquad
  \hat{X} = \tfrac{1}{2}(a + a^\dagger),
  \label{eq:Pleft}
\end{equation}
where the projector $\Pi_{X<0}$ projects onto the negative-position
half-line: it is the quantum counterpart of asking whether a classical
trajectory has ended at $x<0$.  In the position representation it is
$\int_{-\infty}^{0}\ket{x}\bra{x}\,dx$; on the truncated Fock grid we
construct it by diagonalizing the quadrature operator $\hat{X}$ and
summing the eigenprojectors belonging to negative eigenvalues, so that
$P_\mathrm{left}(T)$ is the probability that a measurement of position at
the readout time returns a negative value.  $P_\mathrm{left}$ is
the probability that a measurement of $\hat{X}$ at the final time finds
the oscillator on the left of the dividing surface; it is the quantum
analogue of the classical leak indicator $\mathbf{1}[x(T) < 0]$, with
the values $0$, $1$, and $\tfrac{1}{2}$ interpreted as in the
Introduction.  We use the name \emph{left-half-plane occupation} for
readability, but $P_\mathrm{left}$ is a quadrature-based geometric
proxy for the logical error, not a cat-basis logical error rate; the
distinction is taken up in the Limitations.

\section{Classical Preparation-Space Diagnostics}
\label{sec:classical}

This section builds two classical fields on the preparation disk and asks
what they reveal.  We first define the finite-time sensitivity and the
Lagrangian descriptor, then fix the leak/safe partition by direct
integration and cross-check the transport mechanism against a Melnikov
calculation.  We then show that the sensitivity ridge locates the
transport boundary, and ask when that ridge also predicts, preparation by
preparation, which initial conditions leak.

\subsection{Sensitivity and Lagrangian descriptor fields}

For each grid point, Eqs.~\eqref{eq:cl_x}--\eqref{eq:cl_y} are
integrated together with their linearization: writing the equations as
$\dot{\bm{s}} = \bm{f}(\bm{s}, t)$ with $\bm{s} = (x, y)$, the
$2 \times 2$ matrix $V(t)$ solving the \emph{variational equations}
\begin{equation}
  \dot{V} = J(t)\,V, \qquad V(0) = I_2,
  \qquad J_{ij}(t) = \frac{\partial f_i}{\partial s_j}
  \Big|_{\bm{s}(t)},
  \label{eq:variational}
\end{equation}
is exactly the Jacobian $D\Phi^t$ of the trajectory map along the trajectory.
The finite-time sensitivity is its spectral norm
$S_T = \norm{V(T)}_2$ (largest singular value); the forward Lagrangian
descriptor is the trajectory arc length
$L_T = \int_0^T \norm{\bm{f}(\bm{s}(t), t)}\,dt$, computed by appending
one state variable $\dot L = \norm{\bm{f}}$, $L(0) = 0$, to the system and
integrating it alongside the trajectory.  All integrations use the LSODA solver~\cite{Hindmarsh1983,Petzold1983} with
relative tolerance $10^{-8}$ and absolute tolerance $10^{-10}$.

\subsection{The leak/safe partition and the Melnikov cross-check}
\label{sec:partition}

\begin{figure}[htbp]
\centering
\includegraphics[width=\linewidth]{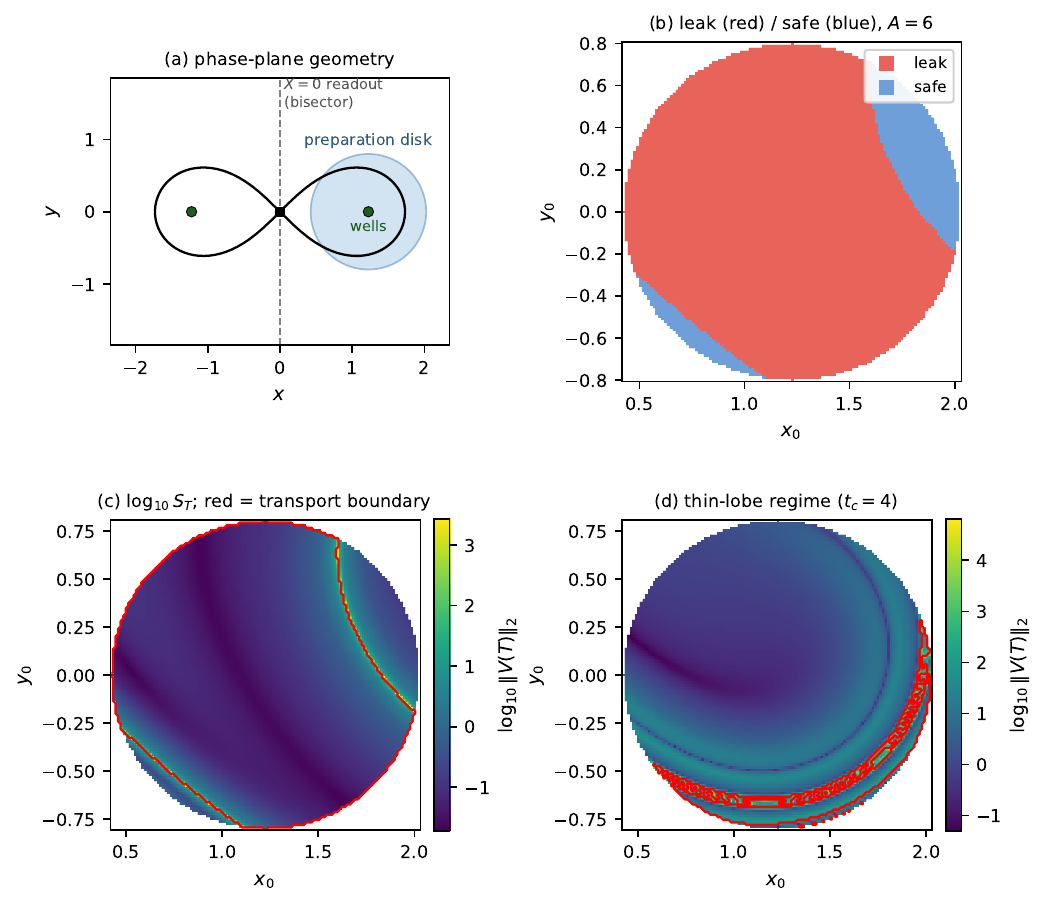}
\caption{\textbf{Classical preparation-space diagnostics} at the
  reference pulse ($A = 6$, $\sigma = 0.3$, $K = \kappa = 1$,
  $p_0 = 1.5$, $T = 8$, $t_c = 0$).
  (a)~Global phase-plane context: the unperturbed ($\kappa = 0$)
  separatrix loops of Eq.~\eqref{eq:separatrix} (black), the saddle at
  the origin, the two wells at $(\pm\sqrt{p_0/K}, 0)$, the dividing
  surface $x = 0$ (dashed), and the preparation disk (shaded).
  (b)~Leak/safe partition of the disk: each preparation is integrated
  to $T$ and colored red if $x(T) < 0$ (it crossed to the left well)
  and blue otherwise.  Leak fraction $= 0.88$.
  (c)~The finite-time sensitivity $\log_{10}\norm{D\Phi^T}_2$ over the
  same disk, with the leak/safe boundary in red; the high-sensitivity
  ridge coincides with the boundary.
  (d)~The thin-crescent regime (late pulse, $t_c = 4$): the leaked set
  is a thin layer coinciding with the high-sensitivity rim, where the
  sensitivity classifies the outcome accurately, in contrast to the
  large-bulk reference case of panel~(c).}
\label{fig:classical}
\vspace{10pt}
\end{figure}

Figure~\ref{fig:classical}(b) shows the partition of the preparation
disk into leaked (red) and safe (blue) initial conditions under the
reference pulse $A = 6$, $\sigma = 0.3$, $t_c = 0$ (leak fraction
$0.88$).  Two independent checks tie the partition to Ref.~\cite{WigKC}.
First, repeating the computation at the $A = 7.5$ pulse analyzed in
Appendix~B.2 of that work reproduces its 150-point preparation ring:
$59$ of the $150$ ring points leak, and they lie where the disk partition
says they should.  Second, the Melnikov function is computed
independently: the integral of the pulse perturbation along the
separatrix Eq.~\eqref{eq:separatrix}.  The dissipative term of the
Melnikov function contains the area enclosed by the separatrix loop,
$p_0/K$, and all Melnikov-derived quantities in the present paper use this
value.

The role of the Melnikov calculation deserves a careful statement,
because its output lives in a different part of the phase plane from
the preparation disk.  The calculation is a global, first-order
cross-check on the turnstile mechanism, not a source of the partition; because $A$ is several times the threshold $A_c = 1.40$, it is used qualitatively (to confirm the turnstile mechanism and locate the intersection points) rather than as a quantitative predictor of the partition.
Its simple zeros in the shift parameter $t_0$ identify intersection
points \emph{on the unperturbed separatrix} at which, to first order, the
pulsed stable and unstable manifolds cross and the turnstile lobe
attaches; these are not preparations and lie outside the preparation
disk.  Fig.~\ref{fig:classical}(a) shows the separatrix geometry and the
disk for scale.  The Melnikov calculation thus confirms the global lobe
mechanism underlying the numerically observed leak/safe partition, while
the partition itself is obtained by direct integration of the pulsed
classical equations.

It is worth emphasizing what the pulse does to the partition: under the
unperturbed flow ($A = 0$) the leaked set is the part of the disk lying
beyond the static basin boundary (about $15\%$ of the disk, on its outer
edge); the pulse replaces this with a different, larger leaked set
anchored at the saddle side.  The partition being predicted is therefore
genuinely pulse-induced transport, not the static basin geometry.

\subsection{The sensitivity ridge coincides with the transport boundary}
\label{sec:ridge}

Figure~\ref{fig:classical}(c) shows $\log_{10} S_T$ over the disk with
the leak/safe boundary overlaid.  The sensitivity field is small almost
everywhere (the dissipative flow contracts) but exhibits a sharp
ridge, and the ridge lies on the boundary.  To quantify this we define
\emph{boundary cells} as grid cells inside the disk having at least one
of their four nearest neighbors with the opposite leak label, and we
rank every cell by its sensitivity percentile (the fraction of cells
with smaller $S_T$).  Boundary cells have a mean sensitivity percentile
of $0.99$; interior cells average $0.49$, as they must by construction.
Conversely, of the top $10\%$ most sensitive cells, the median distance
to the nearest boundary cell is one grid spacing, and $91\%$ lie within
three grid spacings.

The coincidence is robust.  Across pulse amplitudes $A \in \{2, 7.5\}$,
pulse center times $t_c \in \{0, 4\}$, and the no-pulse control, the
mean boundary-cell percentile stays between $0.96$ and $0.99$.  The one
genuine requirement is the length of the integration window: at $T = 2$
the percentile is only $0.55$, recovering to $\approx 0.98$ by $T = 8$.
The reason is dynamical: in a dissipative flow the bulk of either basin
contracts toward an attractor, while trajectories near the stable manifold
of the saddle linger and stretch.  The integration window must be long
enough for the dissipative contraction to separate these two populations;
once it has, $S_T$ localizes onto the boundary.  Lagrangian-descriptor
fields require the same minimum window to develop ridges.

\subsection{When does sensitivity predict the outcome?}
\label{sec:outcome_pred}

Locating the boundary is not the same as predicting, preparation by
preparation, which initial conditions leak, and the reason is geometric.
The sensitivity field is a \emph{ridge}: high \emph{on} the leak/safe
boundary and low on \emph{both} sides of it.  A field that classifies
membership must instead take different values on the two sides, so that a
threshold separates the classes; thresholding the sensitivity selects the
boundary, not the leaked set.  The consequence depends on the shape of the
leaked set, and panels (c) and~(d) of Fig.~\ref{fig:classical} show the
two extremes.

Under the strong early pulse [panel~(c)] the leaked set is a large
interior region, $88\%$ of the disk.  Its bulk has contracted toward the
left attractor and has \emph{low} sensitivity; only its boundary is the
ridge.  The Pearson correlation between $\log_{10} S_T$ and the binary
leak label is therefore $r = -0.45$: the field \emph{anti}-predicts
membership, because most leaked cells are low-sensitivity bulk.  Put as a
ranking, a random leaking preparation outranks a random safe one in
sensitivity only $12\%$ of the time.  The forward Lagrangian-descriptor
field, computed on the same trajectories, behaves the same way.  Under the
late pulse [panel~(d)] the leaked set is a thin crescent hugging the
boundary (leak fraction $0.09$); now the leaked set and the
high-sensitivity boundary nearly coincide, and the ranking reverses, a
random leaking preparation outranking a random safe one $89\%$ of the
time.

The general statement is therefore that the classical sensitivity ridge is
a robust \emph{boundary locator}, and an \emph{outcome predictor} only
when the leaked set is itself a thin layer along the boundary.
Section~\ref{sec:regimes} makes this quantitative as a function of pulse
amplitude and width.

\section{Quantum Preparation-Space Dynamics}
\label{sec:quantum}

This section computes the quantum outcome over the same disk and tests its
robustness.  We first establish the free-evolution baseline, in which the
architecture suppresses transport (Sec.~\ref{sec:static}), then track the
pulsed evolution and show that the quantum left-occupation field follows
the classical transport pattern (Sec.~\ref{sec:scrambling}).  We check
that this picture is not an artifact of strong damping by varying the
single-photon loss rate (Sec.~\ref{sec:kappa}) and then add the engineered
two-photon loss channel (Sec.~\ref{sec:twophoton}).  Finally we form the
gate-induced vulnerability field and show that the engineered stabilizer
recovers the logical bit (Sec.~\ref{sec:dpleft}).

\subsection{Free evolution: transport is suppressed}
\label{sec:static}

Under the constant pump ($A = 0$), the classical flow sends about $15\%$
of the preparation disk into the left basin.  The quantum evolution does
not reproduce this.  Propagating every grid preparation under
Eq.~\eqref{eq:lindblad} with $p(t) = p_0$, the final mean position
$\Re\expect{a}(T)$ never becomes negative, and
$P_\mathrm{left}(T)$ spans $[0.15, 0.43]$ and never exceeds
$\tfrac{1}{2}$: no preparation ends with majority weight in the left
well.  The mechanism is the one that makes the architecture attractive
in the first place: the pump and the dissipation jointly restabilize
each preparation within its own well, and the spontaneous quantum
bit-flip is suppressed on this timescale.  (The wavepacket does
spread (hence $P_\mathrm{left}$ values well above zero) but the
distribution remains right-dominated for every preparation.)

\subsection{Pulsed evolution: the quantum field tracks classical transport}
\label{sec:scrambling}

\begin{figure}[htbp]
\centering
\includegraphics[width=\linewidth]{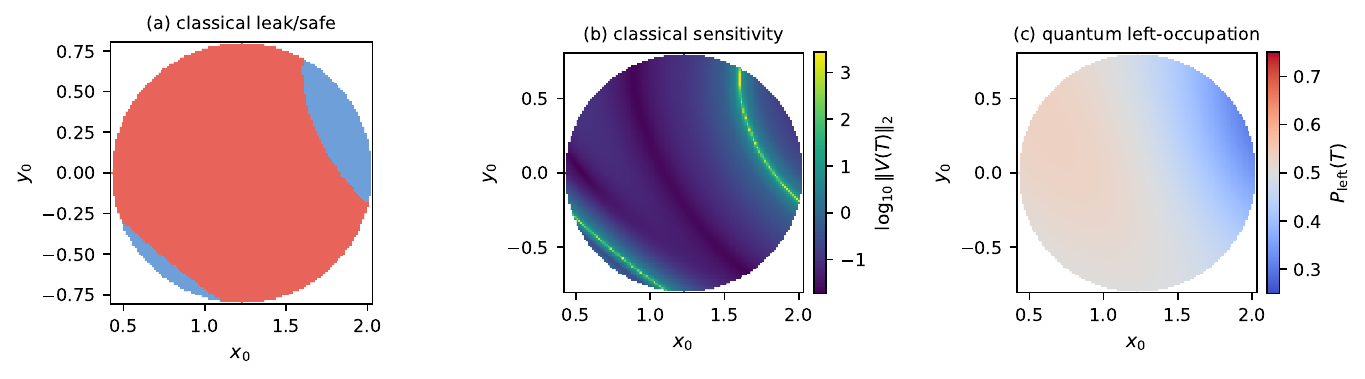}
\caption{\textbf{Classical and quantum preparation-space fields} at the
  reference pulse ($A = 6$, $\sigma = 0.3$, $t_c = 0$, $\kappa = K = 1$,
  $T = 8$, Fock truncation $N = 24$).  (a)~Classical leak/safe partition
  (red = leaked, blue = safe; $88\%$ leaked).  (b)~Classical finite-time
  sensitivity $\log_{10}\norm{D\Phi^T}_2$, whose ridge lies on the
  leak/safe boundary.  (c)~Quantum left-half-plane occupation
  $P_\mathrm{left}(T) \in [0.31, 0.54]$, with the classical boundary
  overlaid in red: the quantum field tracks the classical partition
  ($r = +0.61$) at a smeared, near-even level.}
\label{fig:quantum}
\vspace{10pt}
\end{figure}

Figure~\ref{fig:quantum} compares the classical and quantum fields under
the reference pulse ($A = 6$, $\sigma = 0.3$, $t_c = 0$).  Classically,
$88\%$ of the disk crosses the dividing surface, deterministically and
along a sharp partition.  The quantum outcome differs in degree, not in
spatial organization.

First, $P_\mathrm{left}(T)$ spans only $[0.31, 0.54]$: every preparation
ends near an even probability split between the wells, and none ends
left-localized.  A value $P_\mathrm{left} \approx \tfrac{1}{2}$ means the
final state has its $\hat{X}$-distribution spread almost symmetrically
across both wells, which for our open system is a near-even mixture of the
two well populations.  The pulse does not carry the state cleanly across the
barrier; it delocalizes which-well population over both sides.  Whether the
\emph{logical bit} survives this delocalization is a separate question,
settled by the trace-distance measurement of Sec.~\ref{sec:protocol}.

Second, the spatial structure of the quantum field tracks the classical
partition: the correlation between the classical leak label and
$P_\mathrm{left}$ over the disk is $r = +0.61$.  The quantum
left-occupation is a smeared but faithful image of the classical transport
pattern, in that the preparations the classical model sends across the
dividing surface are those with the largest left-weight, even though no
preparation reaches full transfer.  This positive correspondence is the
basis of the gate-induced vulnerability field of Sec.~\ref{sec:dpleft}.
That the field tracks transport does not mean the bit survives: the
spatial pattern of $P_\mathrm{left}$ is a preparation-resolved field,
whereas whether the two logical inputs stay distinguishable is the
separate question settled in Sec.~\ref{sec:protocol}.  The two can
disagree, and here they do.

\subsection{Robustness to the single-photon loss rate}
\label{sec:kappa}

One might suspect the near-even split is an artifact of strong damping,
since $\kappa = K$ is far from the weakly-dissipative regime in which
the Melnikov construction is formally derived.  It is not.  Reducing
$\kappa$ from $1.0$ to $0.5$ to $0.2$ with all other parameters fixed,
the spread of $P_\mathrm{left}$ over the preparation disk
\emph{narrows} and its values remain pinned near $\tfrac{1}{2}$ (the
maximum stays $\approx 0.54$).  Over this range, weaker dissipation makes the delocalization
more uniform, not less; the delocalization itself is driven by the
coherent two-photon pulse rather than by the loss channel.

\subsection{Robustness to two-photon (pair) loss}
\label{sec:twophoton}

\begin{figure}[htbp]
\centering
\includegraphics[width=\linewidth]{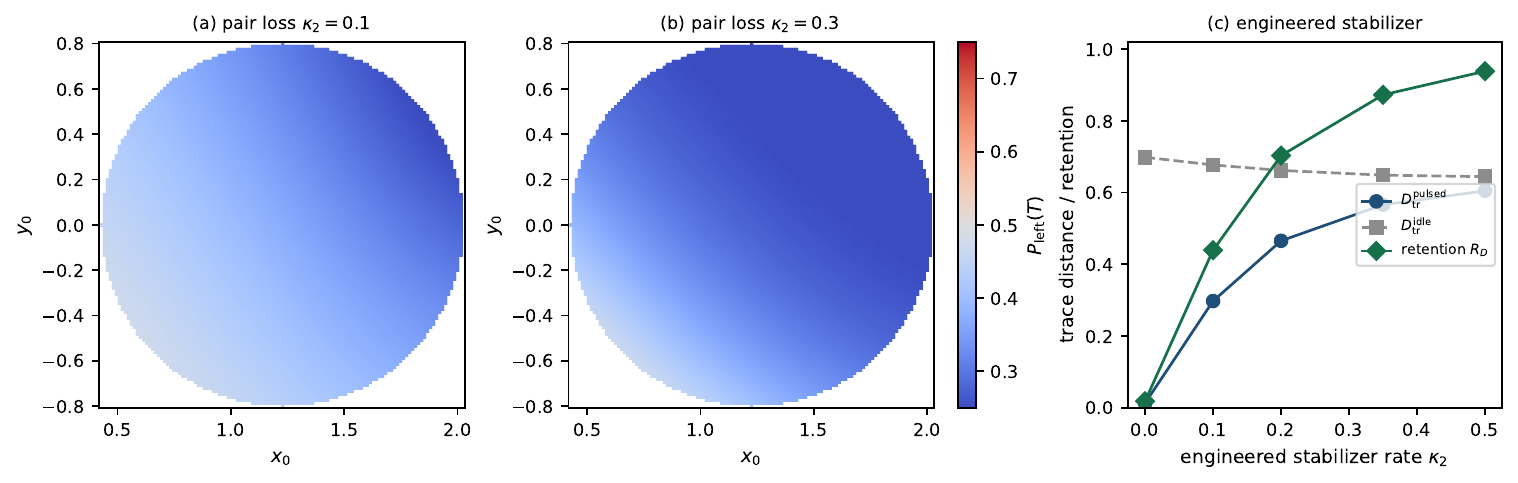}
\caption{\textbf{Two-photon pair loss and the engineered stabilizer}
  ($A = 6$, $\sigma = 0.3$, $\kappa = K = 1$, $T = 8$).
  (a),(b)~Quantum left-half-plane occupation $P_\mathrm{left}(T)$ over the
  preparation disk under phenomenological pair loss at $\kappa_2 = 0.1$ and
  $0.3$.  The classical leak fraction is exactly zero in both cases: the
  classical model predicts no errors, while up to $47\%$ of the which-well
  weight is still lost for the worst preparations, and the Loschmidt echo
  at $T = 1$ remains a strong outcome predictor ($r = -0.96$ and $-0.87$).
  (c)~The engineered cat stabilizer
  $\kappa_2 \mathcal{D}[a^2 - \alpha_\mathrm{cat}^2]$: pulsed and idle
  trace distances and the gate-induced retention
  $R_D = D_{\mathrm{tr}}^{\mathrm{pulsed}}/D_{\mathrm{tr}}^{\mathrm{idle}}$
  versus stabilizer rate $\kappa_2$.  $R_D$ rises from $0.02$ to $0.94$ as
  the stabilizer nearly eliminates the gate-induced loss, while the idle
  distinguishability is only weakly affected.}
\label{fig:twophoton}
\vspace{10pt}
\end{figure}

Physical devices add an engineered two-photon channel; we model it as
the pair-loss dissipator of Eq.~\eqref{eq:lindblad2}
(Sec.~\ref{sec:qmodel}).  We repeated the full computation (quantum grid, classical grid with the corresponding
mean-field damping term, and the Loschmidt echo of Sec.~\ref{sec:loop}) at
$\kappa_2 = 0.1$ and
$\kappa_2 = 0.3$ with the reference pulse.  Three things happen
(Fig.~\ref{fig:twophoton}).

First, classical transport is eliminated: the leak fraction drops from
$0.88$ to exactly $0$ at both $\kappa_2$ values.  The mean-field damping
$-\kappa_2 |\alpha|^2 \alpha$ grows with the excursion amplitude and
arrests precisely the large excursions through which the pulse carried
trajectories across the dividing surface; since the preparation disk reaches no closer than $x_0 \approx 0.42$ to that surface, a crossing requires a large excursion in the first place.

Second, quantum delocalization (the spreading of which-well population across both wells in phase space) is suppressed but not eliminated:
$P_\mathrm{left}(T)$ spans $[0.25, 0.47]$ at $\kappa_2 = 0.1$ and
$[0.23, 0.47]$ at $\kappa_2 = 0.3$, down from $[0.31, 0.54]$.  The
pair-loss channel does its stabilizing job, yet the most vulnerable
preparations (those with the largest $P_\mathrm{left}$) still lose up to $47\%$ of their which-well population to the
far well, a source of logical error that is now entirely invisible to the
classical model, which predicts zero transport.  The classical leak set never predicts the \emph{amount} of logical-bit
loss.  Without pair loss the classical picture overpredicts: it shows
deterministic bit-flips where the quantum system merely scrambles, but it
still ranks correctly which preparations are the most vulnerable.  With
pair loss it underpredicts: it shows no transport at all, while the quantum
system still loses up to $47\%$ of the which-well population for the worst
preparations.

Third, the Loschmidt-echo prediction of Sec.~\ref{sec:loop} survives
intact: the correlation between the echo at $T = 1$ and
$P_\mathrm{left}(T = 8)$ is $r = -0.96$ at $\kappa_2 = 0.1$ and
$r = -0.87$ at $\kappa_2 = 0.3$.
Truncation remains converged (the population in the highest retained Fock levels is $< 10^{-17}$).

\subsection{The gate-induced vulnerability field and an engineered stabilizer}
\label{sec:dpleft}
With the disk centered on the right well and the readout the perpendicular
bisector of the two wells (Sec.~\ref{sec:model}), the left-occupation
tracks classical transport directly: $r(P_{\mathrm{left}}, \mathrm{leak})
= +0.61$ under the quench.  The raw field nonetheless carries a static
coherent-state tail (up to $0.41$ under idle evolution) unrelated to the
gate.  Subtracting it gives the gate-induced field
$\Delta P_{\mathrm{left}} = P_{\mathrm{left}}^{\mathrm{pulsed}}(T)
- P_{\mathrm{left}}^{\mathrm{static}}(T)$, which isolates the pulse's
effect ($r(\Delta P_{\mathrm{left}}, \mathrm{leak}) = +0.53$) and is the
sharper measure of which preparations the pulse makes vulnerable.  Under
the smooth pulse of Sec.~\ref{sec:protocol} both correlations fall
($+0.18$ and $+0.06$).

The simulations above use the phenomenological pair-loss channel
$\kappa_2 \mathcal{D}[a^2]$.  The engineered cat stabilizer is instead
$\kappa_2 \mathcal{D}[a^2 - \alpha_{\mathrm{cat}}^2]$, whose steady states
$\ket{\pm\alpha_{\mathrm{cat}}}$ are the wells; we simulate it explicitly.
Repeating the quench with the stabilizer recovers the logical bit: the
trace distance of Sec.~\ref{sec:protocol} rises from $0.013$ to $0.466$
at $\kappa_2 = 0.2$ and to $0.605$ at $\kappa_2 = 0.5$.  The cleanest
gate-induced measure normalizes against idle evolution under the same
stabilizer; the retention
$R_D(\kappa_2) = D_{\mathrm{tr}}^{\mathrm{pulsed}}/D_{\mathrm{tr}}^{\mathrm{idle}}$
rises from $0.02$ (no stabilizer) to $0.70$ and $0.94$ as
$\kappa_2 = 0.2, 0.5$: the stabilizer \emph{nearly eliminates} the
gate-induced logical loss.  The readout trace distance does not reach
unity ($0.605$), but the shortfall is mostly idle rather than gate-induced:
single-photon loss alone erodes the idle distinguishability to
$D_{\mathrm{tr}}^{\mathrm{idle}} \approx 0.64$ over $T = 8$
[Fig.~\ref{fig:twophoton}(c)].  At $T = 8$, then, a small gate-induced quantum loss remains that the
mean-field model cannot see: that model predicts no crossing, and because
it tracks only the centroid $\expect{a}$ rather than the full state, it has
no trace distance to report.  Whether this residual persists at long times
or is a finite-time effect at $T = 8$ we leave open.

\section{Protocol Dependence: Quench and Smooth Pulses}
\label{sec:protocol}
The reference pulse of the preceding section begins integration at the
center of the Gaussian, so the oscillator experiences a sudden turn-on
of the drive to $p_0 + A$ followed by a decay, a quench.  A physical
gate ramps the drive up and down smoothly.  We now show that this
distinction controls whether the logical bit is corrupted, and we
measure the corruption directly with the trace distance between logical
inputs.  This is the paper's \emph{logical-bit} level:
$D_{\mathrm{tr}}$ is not a field on the preparation disk but the
distinguishability of the two encoded inputs, and it can diverge from the
preparation-resolved fields of Sec.~\ref{sec:quantum}.  It measures
preservation of the encoded which-well bit, not the full quantum channel.

\subsection{Logical information loss, measured directly}
The left-occupation reaching $\tfrac{1}{2}$ shows balanced which-well
weight, but not, by itself, loss of the logical bit: a coherent cat, a
structured state, and an incoherent mixture all give $\tfrac{1}{2}$.  To
measure logical-information loss we prepare the two logical inputs
$\ket{\pm\alpha_{\mathrm{cat}}}$, $\alpha_{\mathrm{cat}} = \sqrt{p_0/K}$,
evolve both under the same pulse, and compute the trace distance
$D_{\mathrm{tr}}(\rho_R, \rho_L) = \tfrac{1}{2}\norm{\rho_R - \rho_L}_1$,
the optimal distinguishability of the two inputs~\cite{Helstrom1969}.  We
use \emph{distinguishability} throughout to mean this trace distance:
$D_{\mathrm{tr}} = 1$ is perfect distinguishability and $D_{\mathrm{tr}} = 0$
a coin flip, the optimal single-shot success probability being
$(1 + D_{\mathrm{tr}})/2$.  The distance is taken between the \emph{full}
evolved states $\rho_R$ and $\rho_L$, not their projections onto the
logical subspace; the subspace projection used below is a separate check on
whether the residual is a coherent superposition or an incoherent mixture.

Under the quench, $D_{\mathrm{tr}}$ collapses from $0.999$ initially to
$0.013$ at readout; the best single-shot discrimination is then
$\approx 51\%$, a coin flip.  Projecting $\rho_R$
onto the orthonormalized logical subspace confirms a near-even
\emph{incoherent} mixture rather than a hidden coherent superposition: the
subspace holds $92\%$ of the population, split $0.494/0.506$ (diagonal
imbalance $0.012$) with off-diagonal coherence only $0.063$.  Idle
evolution leaves the inputs distinguishable ($D_{\mathrm{tr}} = 0.70$), so
the collapse is gate-induced.  This is logical information loss in the
strict sense: the which-well information is erased, not merely spread.

\subsection{A smooth pulse substantially suppresses the loss}
\begin{figure}[htbp]
\centering
\includegraphics[width=\linewidth]{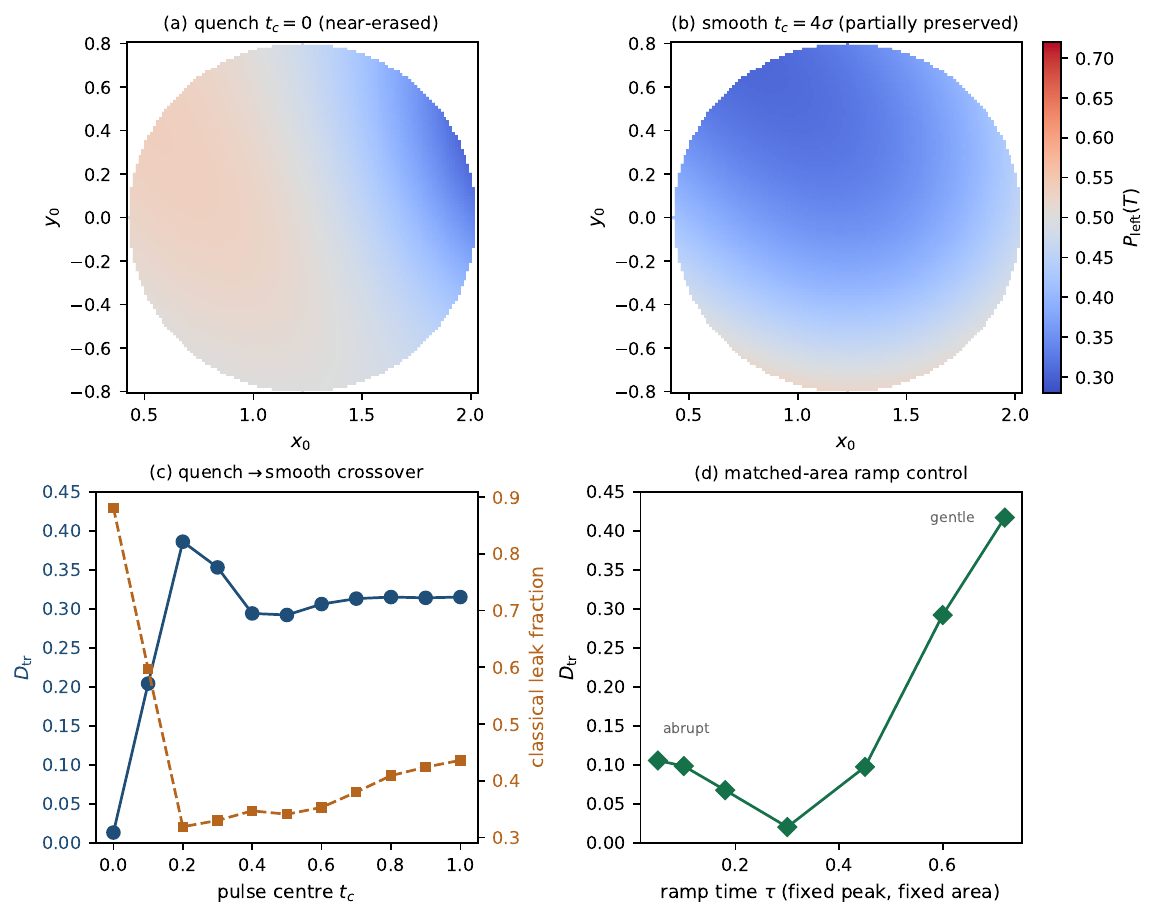}
\caption{\textbf{Protocol dependence.}  (a)~Left-occupation
  $P_{\mathrm{left}}(T)$ over the preparation disk under the quench
  ($t_c = 0$): the field sits near $\tfrac{1}{2}$ and the bit is
  near-erased.  (b)~The same under a full smooth Gaussian with the same peak amplitude
  ($t_c = 4\sigma$): the field stays right-localized and the bit is
  partially preserved; panels share the color scale, centered at
  $\tfrac{1}{2}$.  (c)~Crossover as the pulse center $t_c$ sweeps from
  quench to smooth: the logical trace distance $D_{\mathrm{tr}}$ (left
  axis) rises sharply, overshoots near $t_c \approx 0.2$, and settles to
  $\approx 0.32$, while the classical leak fraction (right axis) drops
  steeply and then drifts up.
  (d)~Matched-area ramp control: logical trace distance versus ramp time
  $\tau$ of a trapezoidal drive at fixed peak and fixed integrated area.
  The dependence is non-monotonic (the gentlest ramp preserves the bit best and the most abrupt is intermediate) so smoothness is not a single
  causal switching-rate variable.}
\label{fig:protocol}
\vspace{10pt}
\end{figure}

Repeating with a smooth pulse centered at $t_c = 4\sigma$, so that the
oscillator experiences the full rise and fall, suppresses the loss across
every diagnostic (Fig.~\ref{fig:protocol}).  The classical leak fraction
falls from $0.88$ to $0.46$; the left-occupation remains predominantly
right-biased (range $[0.31, 0.52]$, only slightly exceeding $\tfrac{1}{2}$
for the most vulnerable preparations); and the bit is largely retained,
$D_{\mathrm{tr}} = 0.32$.  Relative to idle evolution the retention is
$R_D = D_{\mathrm{tr}}^{\mathrm{pulse}}/D_{\mathrm{tr}}^{\mathrm{idle}}
= 0.46$ for the smooth pulse against $0.02$ for the quench, substantial
suppression, not error-free preservation.  The mechanism is consistent
with quasi-adiabatic following: a sudden jump moves the well bottom from
$\sqrt{p_0/K}$ to $\sqrt{(p_0+A)/K}$ and launches large excursions,
whereas a smooth ramp lets the state follow the moving well.  The two
protocols share peak amplitude and width but not integrated area (the quench applies only the decaying half of the Gaussian, the smooth pulse the full profile) so the contrast we report is between protocols, not
between differentiable and non-differentiable drives alone.

To ask whether smoothness \emph{alone} is the operative variable, we ran
a narrow control that isolates the ramp rate by holding peak and integrated
area fixed: a trapezoidal drive of fixed peak and fixed integrated area,
sweeping the ramp time $\tau$ while adjusting the plateau to hold the area.
This pins the variable that the quench-versus-smooth contrast above leaves
confounded, since those two protocols differ in integrated area as well as
in ramp rate.  The dependence of $D_{\mathrm{tr}}$ on $\tau$ is non-monotonic
[Fig.~\ref{fig:protocol}(d)]: the gentlest (triangular) member preserves
the bit best ($D_{\mathrm{tr}} = 0.42$) and the most abrupt (near-top-hat)
member is intermediate ($0.11$), with the worst case between them.  The
sampled dependence is consistent with a competition between the ramp time
and the dwell time at peak drive (period $\approx 0.6$), which co-vary once
peak and area are held fixed; the seven points establish the
non-monotonicity but not a resolved resonance.  Smoothness therefore helps
at the adiabatic end, but a single switching-rate criterion independent of
pulse area is not supported; the quench-versus-smooth distinction is
genuine but multi-parametric.

\subsection{The crossover}
Scanning the pulse center $t_c$ from quench ($0$) to smooth ($4\sigma$),
$D_{\mathrm{tr}}$ rises sharply and \emph{non-monotonically}
(Fig.~\ref{fig:protocol}c): it climbs from $0.013$, overshoots to $0.39$
near $t_c \approx 0.2$, dips to $\approx 0.29$, and settles onto a plateau
near $0.32$.  The classical leak fraction drops steeply from $0.88$ to
$\approx 0.32$ over the same interval and then drifts upward.  The
modulation in the leak fraction is consistent with a response near twice
the small-oscillation frequency at peak drive
($T_{\mathrm{well}} = 2\pi/(\sqrt{2}\,p)$), though the present scan
resolves only a few extrema and we do not fit the frequency.  The quantum
$D_{\mathrm{tr}}$ is the smoother of the two, a further instance of the
wavepacket averaging over fine classical structure.

\subsection{The Loschmidt echo as a quench-regime early-warning diagnostic}
\label{sec:loop}
Let $F(t)$ be the Uhlmann fidelity~\cite{Uhlmann1976,Jozsa1994} between
the pulsed and idling states (the idling state is the qubit held under the
constant pump $p_0$ with no gate) at
time $t$, both evolved forward, and $D_F = -\log F$ the corresponding
disturbance.  Evaluated near the pulse end $t_{\mathrm{eval}}$, $D_F$
predicts the readout outcome in the quench regime:
$r(F, P_{\mathrm{left}}(T)) = -0.94$ at $t_{\mathrm{eval}} \approx 1$.
Despite the name, this echo involves no backward evolution: both states
are evolved forward, the only difference being whether the gate is applied,
so it is computed in real time as the gate runs.  That is what makes it an
early-warning quantity: evaluated near the gate's end, a comparison of the
gated and idle evolutions of the same preparation anticipates the much
later logical outcome.  The prediction is specific to the quench, since
under the smooth pulse the correlation weakens and across the ramp it is
erratic, so the echo is one of the diagnostics that reveals the protocol
dependence.  It is not a computational shortcut, since it requires the same
pulsed state from which $P_{\mathrm{left}}$ is already computed.  The operator-growth (out-of-time-order) analysis that motivated
this echo is collected in Appendix~\ref{app:otoc}; it does not yield a
clean semiclassical correspondence, and we treat it as a negative result.

\section{Cat-Size Dependence}
\label{sec:catsize}
\begin{figure}[htbp]
\centering
\includegraphics[width=\linewidth]{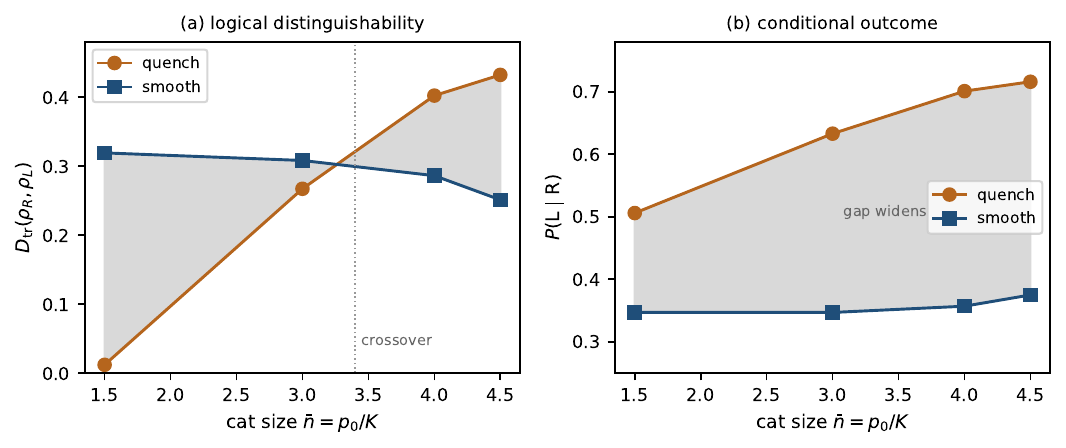}
\caption{\textbf{Cat-size scan} across $\bar n = 1.5, 3.0, 4.0, 4.5$.
  (a)~The logical trace distance $D_{\mathrm{tr}}(\rho_R, \rho_L)$ at
  readout versus cat size $\bar n = p_0/K$, for the quench and the smooth
  pulse.  Under the quench, erasure weakens as the cat grows and
  $D_{\mathrm{tr}}$ rises until it overtakes the smooth-pulse value near
  $\bar n \approx 3.4$; the smooth pulse stays partially preserving.
  (b)~The conditional outcome $P(\mathrm{L}\,|\,\mathrm{R})$: the quench
  drives the right input toward the left well (classical transport) as
  $\bar n$ grows while the smooth pulse keeps it right-localized, so this
  gap widens.}
\label{fig:catsize}
\vspace{10pt}
\end{figure}

The reference cat is small, $\bar n = p_0/K = 1.5$, the mean photon
number of the well coherent state.  This number has a phase-space reading:
$p_0/K$ is the area enclosed by one separatrix loop
(Eq.~\eqref{eq:separatrix}), so the cat size \emph{is} the action enclosed
by a well, and the effective semiclassical parameter $1/\bar n$ is its
inverse; growing the cat enlarges the separatrix loop.  Throughout this
section ``the quench'' is the reference $t_c = 0$ protocol of
Sec.~\ref{sec:protocol}, in which integration begins at the Gaussian peak
and the oscillator feels only the decaying half of the pulse.  Because the
classical--quantum correspondence is governed by the effective
semiclassical parameter $\sim 1/\bar n$, we scan the cat size by scaling
$K \to K_0/s^2$ at fixed $p_0$, which leaves the classical dynamics
invariant under $x \to s\,x$ (the leak partition keeps its shape, leak
fraction $0.88$ at every size) while growing $\bar n = s^2 p_0/K_0$; the
preparation disk, pulse, and Fock truncation are scaled accordingly
(Fig.~\ref{fig:catsize}).

Two quantities are tracked, and keeping them apart is the key to this
scan.  The logical trace distance $D_{\mathrm{tr}}$ asks whether the bit is
\emph{recoverable}, that is, whether the two inputs remain distinguishable;
the conditional outcome $P(\mathrm{L}\,|\,\mathrm{R})$, the probability of
ending in the left well given a right-well input, asks whether the bit has
\emph{flipped}.

Under the quench the erasure weakens as the cat grows, and a weaker erasure
means a larger trace distance: across $\bar n = 1.5, 3.0, 4.0, 4.5$ the
trace distance rises $D_{\mathrm{tr}} = 0.012 \to 0.27 \to 0.40 \to 0.43$
while $P(\mathrm{L}\,|\,\mathrm{R}) = 0.51 \to 0.63 \to 0.70 \to 0.72$
climbs toward a full deterministic flip.  The mechanism is a change in the
kind of failure.  A deterministic swap, a near-deterministic permutation of
the two wells, sends the right well to the left and the left to the right,
so the two inputs land on two \emph{different} outputs that are simply
exchanged; the trace distance is preserved, and the bit, though flipped, is
recoverable.  Erasure instead sends \emph{both} inputs to the \emph{same}
mixed output, so they become indistinguishable and the trace distance
collapses.  As the cat is made more semiclassical the quench moves from
erasure toward a clean swap, which is why the trace distance recovers.

The smooth pulse stays partially preserving at every size tested
($D_{\mathrm{tr}} \approx 0.32, 0.31, 0.29, 0.25$;
$P(\mathrm{L}\,|\,\mathrm{R}) \approx 0.35$).  The two protocols
therefore cross near $\bar n \approx 3.4$: the quench's trace distance, far
below the smooth pulse's at the reference size, overtakes it as the cat
grows.  This is not a reversal of the central finding.  At large cat the
quench preserves \emph{information} better only because a clean swap keeps
the inputs distinguishable; the bit has still flipped, as the widening
$P(\mathrm{L}\,|\,\mathrm{R})$ gap shows.  Two conclusions follow.  The
quench-versus-smooth distinction is robust: it persists and grows with cat
size, so it is not an artifact of the small reference cat.  The near-total
erasure of the bit, by contrast, is specific to the small-cat quench, and
it weakens as the cat grows.

\section{Regime Maps: Pulse Amplitude and Pulse Width}
\label{sec:regimes}

\begin{figure}[htbp]
\centering
\includegraphics[width=\linewidth]{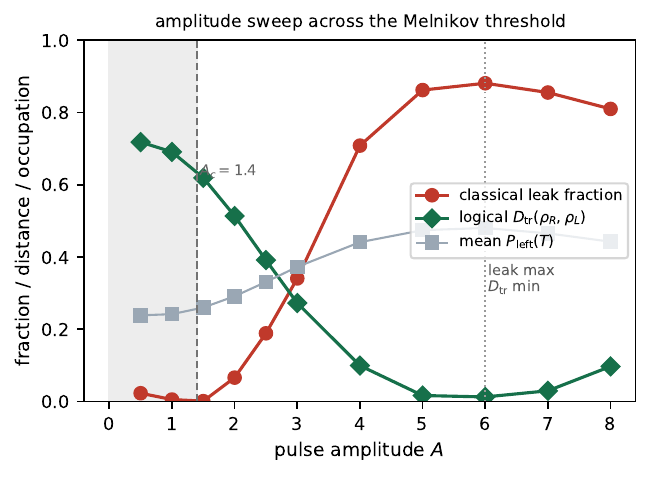}
\caption{\textbf{Amplitude sweep} at $\sigma = 0.3$, $t_c = 0$,
  $\kappa = K = 1$, $T = 8$.  Classical leak fraction, logical trace
  distance $D_{\mathrm{tr}}(\rho_R, \rho_L)$, and quantum mean
  $P_\mathrm{left}(8)$ versus pulse amplitude $A$.  Transport begins near
  $A \approx 2$; the leak-fraction maximum and the $D_{\mathrm{tr}}$
  minimum coincide, at the sampled resolution, near the reference
  amplitude.}
\label{fig:sweep}
\vspace{10pt}
\end{figure}

\begin{figure}[htbp]
\centering
\includegraphics[width=\linewidth]{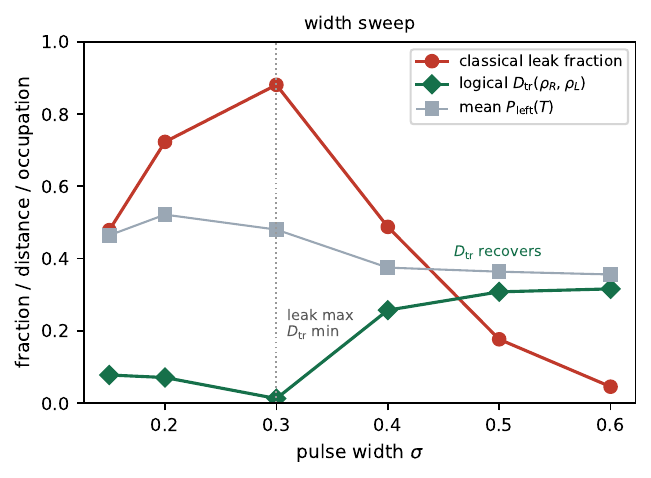}
\caption{\textbf{Width sweep} at $A = 6$, $t_c = 0$, $\kappa = K = 1$,
  $T = 8$.  Classical leak fraction, logical trace distance
  $D_{\mathrm{tr}}$, and quantum mean $P_\mathrm{left}(8)$ versus pulse
  width $\sigma$.  The leak fraction peaks at $\sigma = 0.3$, exactly where
  $D_{\mathrm{tr}}$ is minimal; wider pulses recover $D_{\mathrm{tr}}$
  monotonically as the gate becomes quasi-adiabatic.}
\label{fig:sigma}
\vspace{10pt}
\end{figure}

The results so far are stated at one reference configuration, which we fix
here for the sweeps that follow.  The reference pulse has amplitude
$A = 6$, width $\sigma = 0.3$, and center $t_c = 0$, and the reference cat
has $\bar n = p_0/K = 1.5$ (Secs.~\ref{sec:model} and~\ref{sec:catsize});
by \emph{the reference amplitude} we mean $A = 6$.  This pulse is well
above the first-order Melnikov threshold, deep in the strong-driving
regime.  Throughout this section \emph{distinguishability} is the logical
trace distance $D_{\mathrm{tr}}$ of Sec.~\ref{sec:protocol}, taken between
the full evolved logical inputs.  To locate the boundaries of the
conclusions above, we repeated the joint computation (classical partition,
quantum
$P_\mathrm{left}(8)$, the logical trace distance $D_{\mathrm{tr}}$, and echo
$F(1)$) across pulse amplitude and width
(Figs.~\ref{fig:sweep},~\ref{fig:sigma}).

\emph{Amplitude.}  At fixed $\sigma = 0.3$, the classical leak fraction is
negligible below $A \approx 2$ (the first-order Melnikov threshold is
$A_c = 1.40$), rises steeply through $A = 3$--$4$, and saturates near
$0.88$ at the reference amplitude.  The quantum mean $P_\mathrm{left}$
rises monotonically with $A$ toward the even-split value $\approx 0.48$.
The logical trace distance falls in lockstep with the leak fraction: from
$D_{\mathrm{tr}} \approx 0.7$ below threshold to its \emph{minimum}
($0.013$) near $A \approx 6$, where the classical leak is maximal, then
recovering slightly by $A = 8$.  Because $P_\mathrm{left} \approx 1/2$ is
not by itself erasure (Sec.~\ref{sec:protocol}), it is the trace distance
that makes the coincidence meaningful: the classical-leak maximum and the
logical-distinguishability minimum occur together, at the sampled
resolution, near the reference amplitude.

\emph{Width.}  At fixed $A = 6$, the classical leak fraction peaks sharply
at $\sigma = 0.3$ ($0.88$) and collapses for wider pulses ($0.18$ at
$\sigma = 0.5$, $0.05$ at $\sigma = 0.6$): a wide pulse lets the state
follow the moving well quasi-adiabatically, that is, slowly enough that
it stays in the well as the well shifts, and suppresses transport; the
same mechanism as the smooth protocol of Sec.~\ref{sec:protocol}.  The
quantum mean $P_\mathrm{left}$ falls correspondingly (from $\approx 0.48$
to $\approx 0.36$).  The logical trace distance mirrors the leak fraction
inversely: it is \emph{minimal} ($0.013$) at $\sigma = 0.3$, where the
classical leak is maximal, and recovers monotonically for wider pulses
($D_{\mathrm{tr}} = 0.26, 0.31, 0.32$ at $\sigma = 0.4, 0.5, 0.6$).  At
fixed strong drive it is the width, not the amplitude, that separates the
impulsive (scrambling) and quasi-adiabatic (suppressing) regimes, and the
two sweeps answer the same question consistently: the classical-leak
maximum locates the logical-distinguishability minimum; the rapid decrease
of distinguishability begins in the neighborhood of the classical
Melnikov threshold (with the minimum well past the threshold, at much stronger drive); and increasing width recovers it.

\section{Discussion}
\label{sec:discussion}

\subsection{What each diagnostic measures}

Table~\ref{tab:corr} collects the Pearson correlation for each pair of
the four retained diagnostics at the reference pulse.  The classical sensitivity
locates the transport boundary (its correlation with the leak label,
$r = -0.45$, is negative because the leaked bulk is low-sensitivity,
Sec.~\ref{sec:outcome_pred}).  The classical leak
label \emph{does} track the quantum outcome at this pulse
($r(\mathrm{leak}, P_\mathrm{left}) = +0.61$): the basin and turnstile geometry
predicts the spatial pattern of the quantum left-weight, even though it
cannot capture the magnitude of the loss: the bit erasure that the trace
distance reports.  The Loschmidt echo predicts the outcome directly
($r(F, P_\mathrm{left}) = -0.94$).  The gate-induced OTOC aligns with the
classical sensitivity only \emph{mid-pulse} ($r(\Delta C, S_T) = +0.97$ at
$t_\mathrm{eval} = 0.5$), and the alignment collapses to $+0.09$ by pulse
end; we record it as a negative result (Appendix~\ref{app:otoc}).  No two
diagnostics are interchangeable; each captures one facet.  The regime maps
of Sec.~\ref{sec:regimes} add the qualification that the leak--outcome
correspondence is strongest below and just above the Melnikov threshold
and weakens under saturating strong drive, as the outcome pins toward the
even split.

\begin{table}[h]
\centering
\caption{Pairwise Pearson correlations between the surviving
  preparation-space diagnostics at the reference pulse ($A = 6$,
  $\sigma = 0.3$, $\kappa = K = 1$), on the joint $41 \times 41$ grid.
  Fields are evaluated at $T = 8$ except the echo $F$, taken at the
  pulse-end time $t_\mathrm{eval} = 1$.  The gate-induced OTOC is a
  transient mid-pulse effect and is reported separately in
  Appendix~\ref{app:otoc}.}
  \vspace{20pt}
\label{tab:corr}
\begin{tabular}{lcccc}
\hline\hline
 & Class.\ leak & Class.\ $S_T$ & $P_\mathrm{left}$ & $F(1)$ \\
\hline
Classical leak   & 1.00 & $-0.45$ & $+0.61$ & ---     \\
Classical $S_T$  &      &  1.00   & $-0.58$ & ---     \\
$P_\mathrm{left}$&      &         &  1.00   & $-0.94$ \\
$F(1)$           &      &         &         & 1.00    \\
\hline\hline
\end{tabular}
\end{table}

\subsection{Relation to the double-well study}

The companion paper~\cite{WigJCP} computed the sensitivity field and
the Lagrangian-descriptor field over a preparation space of Gaussian
wavepackets in a driven double well, and found the two fields (both organized by the same phase-space structures) to correlate with each
other at only $r \approx 0.10$: complementary, not interchangeable.
The present results extend that conclusion in three directions.  The
setting is now open (dissipative) rather than conservative, and the
non-interchangeability persists.  The set of computed fields is enlarged: alongside the sensitivity and
the LD of that work, we compute the gate-induced OTOC $\Delta C$, the
echo $F$, and the outcome $P_\mathrm{left}$ itself, and the
complementarity persists: every off-diagonal correlation in
Table~\ref{tab:corr} except the $F$--$P_\mathrm{left}$ pair is well away
from $\pm 1$.  And the mechanism is identified in each case: the classical
sensitivity is a local property of the trajectory map (it lives on the stable
manifold); the OTOC is an operator-growth property that coincides with
the sensitivity only mid-pulse, before the background dynamics
dominate; the echo is a
state-disturbance property that integrates the gate's effect and
therefore tracks the outcome; and the outcome is an open-system
quantity whose spatial pattern the classical transport geometry predicts
($r = +0.61$) but whose magnitude (the erasure) it does not.

\subsection{Relation to the OTOC literature}

Our background OTOC behaves exactly as the $\lambda_L = 0$ result of
Ref.~\cite{AlmasriReboiro} predicts.  With only polynomial commutator
growth, eight Kerr times of background evolution accumulate spatial
OTOC structure that overwhelms the gate's: by $T = 8$ the gate's
contribution, although comparable in magnitude to the background, has
lost its preparation-space structure (spatial standard deviation
$\approx 25\%$ of its mean and uncorrelated with the sensitivity), so
the preparation-space structure of the total OTOC is essentially that
of the background, and the gate's geometric signal is recoverable only
by subtraction and only at short time.  This is the practical face of
the absent exponential-growth window: in a system with a positive quantum
Lyapunov exponent the gate's perturbation would be exponentially amplified
and dominate the OTOC; here it is buried.  In
systems where the OTOC--Lyapunov correspondence has been demonstrated,
such as the Dicke model~\cite{ChavezCarlos2019}, the correspondence is
between growth \emph{rates}; the field-against-field correspondence
tested here ($r = +0.97$ between $\Delta C$ and the classical
sensitivity, mid-pulse at $t_\mathrm{eval} = 0.5$) is the analogue
available when both rates vanish, and, as Sec.~\ref{sec:timedep}
shows, it is correspondingly fragile, collapsing to $r = +0.09$ by pulse
end.  We work at resonance, far from the exceptional points where
Ref.~\cite{AlmasriReboiro} finds enhanced dephasing, and our
conclusions do not bear on that regime.

\subsection{Design implications}
\label{sec:design}

For pulse design on this architecture the results suggest a protocol
with three regimes.  Below the Melnikov threshold
($M_\mathrm{max} < 0$), the classical computation is sufficient for
locating the spatial vulnerability pattern and cheap at every width tested, and empirically remains adequate somewhat
above the threshold, while the first-order lobe is still a small
fraction of the preparation region: the basin/lobe geometry predicts the
spatial structure of the quantum left-weight at $r \approx +0.5$ to
$+0.8$, and the sensitivity ridge marks the preparations to avoid.
Under strong driving the classical geometry still says something
important, but incomplete.  Under the reference quench it predicts the
spatial \emph{ordering} of vulnerability
($r(\mathrm{leak}, P_\mathrm{left}) = +0.61$) but neither the magnitude of
quantum transfer nor the survival of logical distinguishability, which
collapses regardless.  The operative quantity for the latter is the
Loschmidt echo (a gate-end comparison of the gated and idle evolutions of the same preparation) which predicts the final which-well information loss
at $r \approx -0.94$ at the reference quench, robustly under two-photon
pair loss, until the outcome saturates; a pulse that saturates the
scrambling is rejected without any per-preparation prediction.  For slow
(quasi-adiabatic) pulses, classical transport is suppressed but quantum
delocalization persists, the echo is too shallow to read, and the
honest statement is that no computed diagnostic predicts the residual
quantum error well; slow gates are not automatically safe and must be
validated against the full quantum computation.  Two further cautions
follow from the dissipation and width studies.  Two-photon dissipation
changes the \emph{sign} of the classical error, meaning the classical
prediction minus the quantum reality.  Without pair loss the classical
model overpredicts: it predicts a clean deterministic bit-flip while the
quantum system scrambles, so the predicted error exceeds the actual loss of
distinguishability.  With pair loss it underpredicts: it predicts zero
transport while up to $47\%$ of the which-well population is still lost.
Wide pulses produce the same underprediction with no engineered dissipation
involved.  A design
workflow that validates gates against the classical model alone would,
in either situation, certify pulses that in fact delocalize the qubit.

\subsection{Limitations}

The preparation space considered is the coherent-state family; squeezed
or cat-state preparations define larger preparation spaces on which the
diagnostics are not yet compared.  $P_\mathrm{left}$ is the
$\hat{X}$-quadrature half-line probability; a cat-basis readout would
define a different (though related) outcome variable.  The phenomenological
two-photon channel $\kappa_2\mathcal{D}[a^2]$ and the engineered
stabilizer $\kappa_2\mathcal{D}[a^2 - \alpha_\mathrm{cat}^2]$ are both
simulated (Secs.~\ref{sec:twophoton},~\ref{sec:dpleft}) at two rates each;
stronger stabilizer rates and the slow convergence to the stabilized
manifold beyond $T = 8$ are not mapped.  The matched-area ramp control of
Sec.~\ref{sec:protocol} necessarily co-varies the ramp duration and the
dwell time at peak drive, and its seven-point scan does not resolve whether
the intermediate minimum is a robust dynamical feature.  The pulse family
is otherwise the single Gaussian of Eq.~\eqref{eq:pulse}; shaped or
composite pulses are untested, and the sweeps cover
$\sigma \in [0.15, 0.6]$ and amplitude ratios $A/A_c \lesssim 6$ only.  Under strong driving no single
scalar
($A/A_c$, $M_\mathrm{max}$, or leak fraction) organizes all of the
strongly driven phenomenology, and the conditions under which the
echo--outcome correlation is near perfect have been located empirically
(a large gate disturbance acting on an outcome not yet saturated to the even split) but not
characterized predictively; the echo's optimal evaluation time is also
width dependent.  The predictor of choice is therefore regime
dependent, and assigning a new pulse family to its regime currently
requires the classical computation plus a coarse quantum check.  The
quasi-adiabatic interpretation of the wide-pulse branch is supported by
the monotone suppression of the leaked fraction below its static value
but has not been confirmed by an explicit adiabaticity diagnostic
(e.g.\ tracking the instantaneous equilibria of the frozen-time flow).
Finally, all evolutions extend to $T = 8\,K^{-1}$; we stop at $T = 8$, so any slow
relaxation onto the code manifold at later times is not studied.

\section{Conclusion}
\label{sec:conclusion}

We computed a family of classical and quantum preparation-space
fields jointly on the hardware model of a driven Kerr-cat qubit and
compared them on a common preparation space, separating the
preparation-resolved fields from the logical-input trace distance.  The classical finite-time sensitivity
ridge is a robust locator of the transport boundary (boundary cells at the
99th sensitivity percentile across the regimes tested).  The central
finding is that the gate-induced corruption of the logical bit depends
strongly on the full temporal protocol: ramp, duration, dwell, amplitude,
and width.  A sudden quench erases the bit: the trace
distance between the two logical inputs collapses to
$D_{\mathrm{tr}} = 0.013$, and projection onto the logical subspace shows a
near-even incoherent mixture of the wells (subspace population $0.92$,
diagonal imbalance $0.012$).  A full smooth Gaussian with the same peak amplitude suppresses
the loss, retaining about half the idle distinguishability
($R_D = 0.46$).  With the disk centered on the well and the readout the
perpendicular bisector of the two wells, the quantum left-occupation
tracks classical transport ($r = +0.61$ under the quench), and a Loschmidt
echo evaluated near the pulse end predicts the much later outcome
($r = -0.94$), specifically within the quench regime.

The remaining results delimit and stress-test this picture.  A cat-size
scan shows the erasure weakening as the oscillator is made more
semiclassical, the quench shifting toward deterministic classical
transport.  An engineered pair-loss stabilizer drives the classical leak fraction to
zero and raises the gate-induced retention to $R_D = 0.94$, leaving only a
finite-time residual outside the corresponding first-moment mean-field
model.  Amplitude and width sweeps place the reference pulse at the joint
extremum where the classical leak is maximal and the logical
distinguishability minimal.  The operator-growth (out-of-time-order) diagnostic, which we
had hoped would bridge the classical and quantum pictures, tracks the
classical sensitivity only at the earliest evaluation times and loses the
correspondence thereafter; we record it as a negative result.  The
practical message is a warning about protocol implementation: the dramatic
logical erasure under the reference quench is substantially suppressed by
a smoothly ramped gate of the same amplitude.  Abruptness alone, however,
is not the unique cause; a matched-area control shows the protocol
dependence is multi-parametric.


\appendix
\section{Numerical Details}
\label{app:numerics}

\paragraph{Classical variational integration.}
The augmented state $(x, y, m_{11}, m_{12}, m_{21}, m_{22}, L)$ is
integrated with LSODA, tolerances $\mathrm{rtol} = 10^{-8}$,
$\mathrm{atol} = 10^{-10}$, maximum step $\sigma/2$.  The spectral norm
$S_T = \sigma_{\max}(V(T))$ is extracted by singular-value
decomposition.

\paragraph{Lindblad propagation.}
The vectorised Liouvillian is constructed in the Kronecker
representation; consistency with the direct master equation is verified
to $< 5 \times 10^{-16}$ (maximum entrywise error for a random density
matrix).  For the static case ($A = 0$) the propagator
$U(T) = e^{\mathcal{L}T}$ is a single matrix exponential.  For the
pulsed case a piecewise-constant time-stepping scheme is used: during
the pulse window the step is $\min(0.04, \sigma/8)$ and the window
extends to $t_c + 4\sigma$; outside it the step is $0.3$--$0.4$; at
each step $e^{\mathcal{L}(t_\mathrm{mid})\Delta t}$ is accumulated into
a single $N^2 \times N^2$ propagator, which is then applied to all grid
preparations at once.

\paragraph{Heisenberg propagation and adjoint check.}
The adjoint superoperator $\mathcal{L}^\dagger$
[Eq.~\eqref{eq:heisenberg} in vectorised form] is verified by
evaluating $\Tr[A^\dagger \mathcal{L}\rho] -
\Tr[(\mathcal{L}^\dagger A)^\dagger \rho]$ for random $\rho$ and $A$;
the residual is $< 8 \times 10^{-14}$.  The Heisenberg-evolved operator
$\hat{X}_H(T)$ satisfies Hermiticity to $< 6 \times 10^{-15}$.

\paragraph{Loschmidt echo.}
The fidelity Eq.~\eqref{eq:LE} is evaluated per preparation via the
eigendecomposition $\rho_p = V \Lambda V^\dagger$, computing
$\sqrt{\rho_p} = V\Lambda^{1/2}V^\dagger$ (eigenvalues clipped at
zero), then forming $\sqrt{\rho_p}\,\rho_s\,\sqrt{\rho_p}$ and summing
the square roots of its eigenvalues.

\paragraph{Two-photon dissipation runs.}
The dissipator $\kappa_2\mathcal{D}[a^2]$ is added to the vectorised
Liouvillian in the same representation; the mean-field correspondence
($-\kappa_2|\alpha|^2\alpha$ added to $\dot{\alpha}$) is verified by
short-time propagation of a coherent state as in
Sec.~\ref{sec:qmodel}.  The classical comparison runs use the
correspondingly modified vector field and Jacobian.

\paragraph{Amplitude and width sweeps.}
For each pulse the propagator is accumulated once with snapshots at
$t_\mathrm{eval}$ and $T = 8$; the static propagator is pulse
independent and computed once, with
$e^{\mathcal{L}_s \cdot 8} = (e^{\mathcal{L}_s \cdot 1})^{8}$ for the
amplitude sweep.  The Melnikov threshold $A_c(\sigma)$ is obtained by
root finding on $\max_{t_0} M(t_0; A, \sigma)$, which is affine in
$A$, with the separatrix-loop area $p_0/K$ in the dissipative term.
Fock truncation remains converged in all sweep runs (top-level
population $\lesssim 2 \times 10^{-17}$).

\paragraph{Code and data availability.}
The numerical scripts, data, and figure-generation files used for this
study are available from the author upon reasonable request.

\section{Operator Growth and the Out-of-Time-Order Correlator}
\label{app:otoc}
We collect here the gate-induced out-of-time-order correlator (OTOC)
analysis.  We had anticipated that the OTOC would provide a semiclassical
bridge between the quantum sensitivity and the classical tangent map.  It
does not: across evaluation times no single Jacobian component (and not the spectral norm $\norm{V(T)}_2$ used in the main text) tracks the
differential OTOC field consistently, and the baseline-subtracted
quantity $\Delta C$ has no clean classical correspondence.  We record the
analysis as a negative result and the diagnostic itself as an
adjoint-map squared-commutator probe of operator sensitivity.

\label{app:otocdetail}

\subsection{Definitions and numerical verification}

This section introduces the two quantum diagnostics named in the Introduction (the gate-induced OTOC $\Delta C$ and the Loschmidt echo $F$) and asks (Sec.~\ref{sec:timedep}) whether and when each tracks the classical sensitivity field or predicts the quantum outcome.  Throughout, $\rho$ is the density operator of the oscillator [Eq.~\eqref{eq:lindblad}], $\ket{\alpha_0}$ the coherent-state preparation at $\alpha_0 = x_0 + iy_0$, and the \emph{Heisenberg operator} $\hat{A}_H(T)$ is the observable evolved under the adjoint of the Lindblad generator so that $\Tr[\hat{A}\,\rho(T)] = \Tr[\hat{A}_H(T)\,\rho_0]$.

\paragraph{Heisenberg propagation and the gate-induced OTOC.}
The OTOC requires the Heisenberg-evolved operator $\hat{X}_H(T)$.  In an
open system, operators evolve under the \emph{adjoint} of the Lindblad
generator~\cite{Sakurai2021,Breuer2002}: whereas the state obeys $\dot{\rho} = \mathcal{L}(t)\rho$
[Eq.~\eqref{eq:lindblad}], an observable $A$ obeys
\begin{equation}
  \frac{dA}{dt} = \mathcal{L}^\dagger(t) A
  = i[H(t), A]
  + \kappa\Bigl(a^\dagger A a - \tfrac{1}{2}\{a^\dagger a, A\}\Bigr),
  \label{eq:heisenberg}
\end{equation}
the two generators, the Schr\"odinger generator $\mathcal{L}$ (acting on states) and its adjoint $\mathcal{L}^\dagger$ (acting on observables), being related, at each instant, by the duality
$\Tr[A^\dagger \mathcal{L}\rho] =
\Tr[(\mathcal{L}^\dagger A)^\dagger \rho]$.  For the time-dependent
generator here the operator ordering must be respected.  Discretize
$[0, T]$ into steps with step superoperators
$\mathcal{E}_k = \exp(\mathcal{L}(t_k)\,\Delta t)$.  The state propagates
as $\rho(T) = \mathcal{E}_n \circ \cdots \circ \mathcal{E}_1[\rho_0]$
(step $1$ first), whereas the Heisenberg operator is built from the
adjoint steps in the \emph{reverse} order,
$\hat{X}_H(T) = \mathcal{E}_1^\dagger \circ \cdots \circ
\mathcal{E}_n^\dagger[\hat{X}]$.  Composing the adjoint steps in forward
order instead returns the Heisenberg operator of the
\emph{time-reversed} pulse, a distinct object whenever the drive is
time dependent (for a time-independent generator the two orderings
coincide).  We therefore validate the propagation by
an ordering-sensitive check: the agreement of $\Tr[\hat{X}_H(T)\,\rho_0]$
with $\Tr[\hat{X}\,\rho(T)]$, where $\rho(T)$ is obtained from
independent Schr\"odinger-picture propagation of random preparations
$\rho_0$, holds to the propagator's discretization error
($\sim 10^{-6}$) for the time-ordered construction, whereas the
forward-ordered construction violates it by $O(10^{-2})$.  The per-step
adjoint identity above (residual $< 8 \times 10^{-14}$) and the
Hermiticity of $\hat{X}_H(T)$ ($< 6 \times 10^{-15}$) are necessary but
not sufficient: an instantaneous identity at fixed time cannot detect
the ordering.
With $\hat{X}_H(T)$ in hand, the OTOC of the Introduction is evaluated
for every grid preparation, once with the pulse, $C_\mathrm{pulsed}$,
and once under the constant pump, $C_\mathrm{static}$, and the gate-induced OTOC of the Introduction is formed as the difference
$\Delta C = C_\mathrm{pulsed} - C_\mathrm{static}$.

\paragraph{Loschmidt echo.}
The echo is the quantum fidelity between the pulsed and free evolutions
of the same preparation,
\begin{equation}
  F(T, \alpha_0) =
  \Bigl(\Tr\sqrt{\sqrt{\rho_p}\,\rho_s\sqrt{\rho_p}}\Bigr)^{\!2},
  \label{eq:LE}
\end{equation}
with $\rho_{p}$ and $\rho_{s}$ the states at time $T$ evolved from
$\ket{\alpha_0}$ under Eq.~\eqref{eq:lindblad} with and without the
pulse.  The fidelity equals $1$ if and only if the two states coincide
and decreases toward $0$ as they become distinguishable; small $F$ means
the gate strongly disturbed the preparation relative to free evolution.
A remark on the name: for unitary dynamics, the fidelity between two
forward evolutions is equivalent to the original echo construction of
Peres~\cite{Peres1984} (evolve forward under one Hamiltonian, then
backward under a perturbed one), but for open-system dynamics no
backward evolution exists, and $F$ is the Uhlmann fidelity~\cite{Uhlmann1976,Jozsa1994} between two
forward-evolved density operators.  We keep the name Loschmidt echo~\cite{GorinReview2006}
for brevity.
The unpulsed reference is the stable idling configuration of the qubit;
as established in Sec.~\ref{sec:static}, it produces no bit flips of
its own, so a reduced echo measures disturbance attributable
specifically to the gate.

\subsection{Predictive power as a function of evaluation time}
\label{sec:timedep}

\begin{figure}[htbp]
\centering
\includegraphics[width=\linewidth]{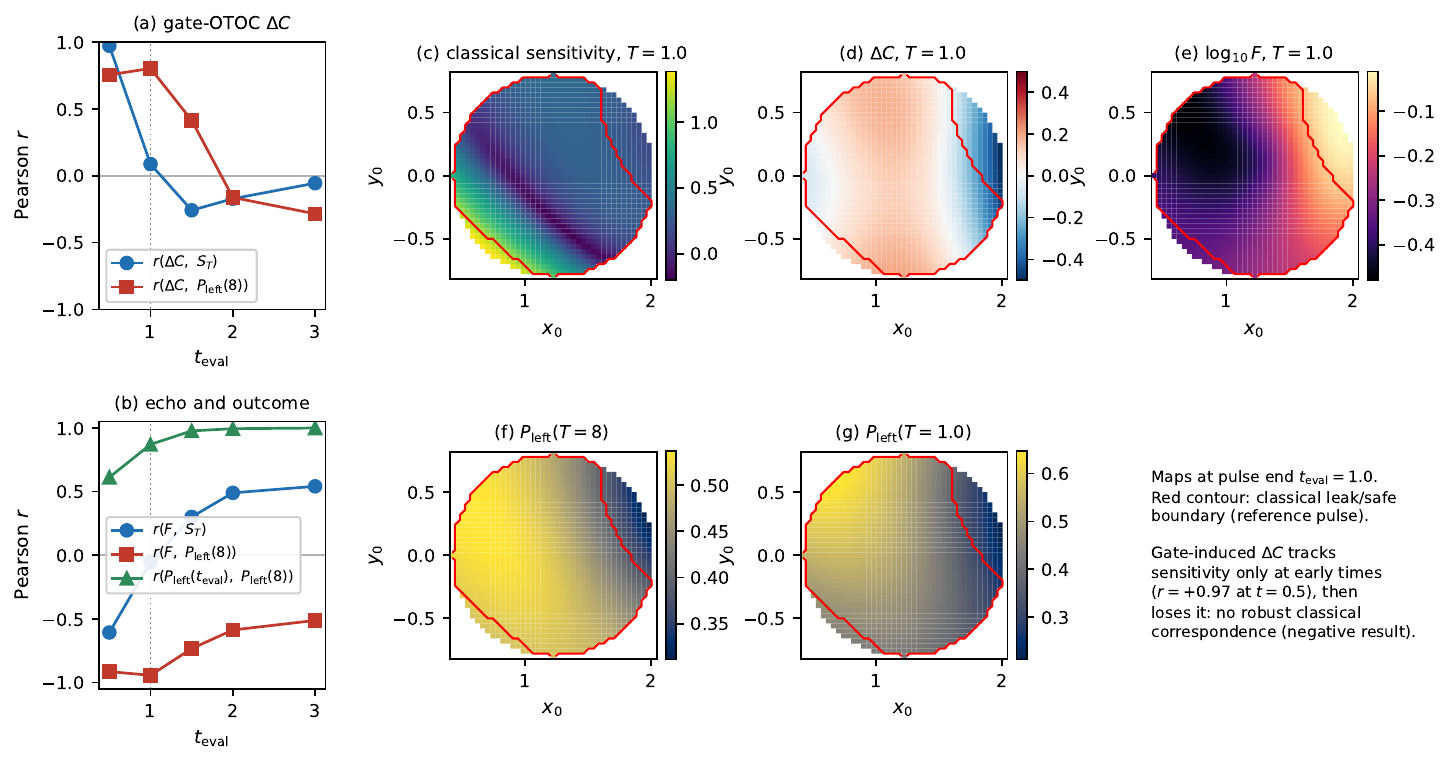}
\caption{\textbf{Time-resolved diagnostics} at the reference pulse
  ($A = 6$, $\sigma = 0.3$, $\kappa = K = 1$).
  (a),(b)~Pearson correlations versus the evaluation time
  $t_\mathrm{eval}$ (dotted vertical line: pulse end, $T \approx 1$).
  In~(a), the time-ordered gate-induced OTOC $\Delta C$ against the
  classical sensitivity computed at the same time (blue) and against
  the final outcome (red): the OTOC--sensitivity correlation is
  positive and strong mid-pulse ($+0.97$ at $t_\mathrm{eval} = 0.5$) but
  collapses by pulse end ($+0.09$ at $t_\mathrm{eval} = 1.0$), where
  $\Delta C$ instead attains its strongest correlation with the outcome
  ($+0.80$).  In~(b), the Loschmidt echo against the final outcome
  (red; $r = -0.91$ at $t_\mathrm{eval} = 0.5$), the echo against the
  classical sensitivity (blue), and the outcome at $t_\mathrm{eval}$
  against the final outcome (green; $0.99$ by $t_\mathrm{eval} = 2$).
  (c)--(g)~The spatial fields at pulse end $t_\mathrm{eval} = 1.0$:
  classical sensitivity, $\Delta C$, $\log_{10}F$, and the outcome at
  $T = 8$ and $T = 1.0$, each with the classical leak boundary in red.
  The gate-induced OTOC~(d) has lost its alignment with the classical
  sensitivity~(c) by pulse end (the collapse quantified in panel~a); the
  echo field~(e) marks the most-disturbed preparations
  (dark = most disturbed) and predicts the final outcome~(f).}
\label{fig:loop}
\vspace{10pt}
\end{figure}

Both $\Delta C$ and $F$ depend on the time $t_\mathrm{eval}$ at which
they are evaluated, and their usefulness turns out to depend on it
strongly.  Figure~\ref{fig:loop} shows their correlations with two
targets, the classical sensitivity field evaluated at the same time,
and the final quantum outcome $P_\mathrm{left}(T = 8)$, for
$t_\mathrm{eval} \in \{0.5, 1.0, 1.5, 2.0, 3.0\}$ at the reference
pulse.

\paragraph{The OTOC--sensitivity bridge is mid-pulse and transient.}
The semiclassical correspondence
$C \sim \hbar^2(\partial x/\partial x_0)^2$ between the gate-induced
OTOC and the classical sensitivity is visible only briefly, and only
while the pulse is still acting.  Mid-pulse, at $t_\mathrm{eval} = 0.5$,
$\Delta C$ and the classical sensitivity field correlate strongly, at
$r = +0.97$, and the two spatial fields are aligned: this is the
correspondence realized field against field.  It does not survive to
pulse end.  By $t_\mathrm{eval} = 1.0$, just as the pulse closes, the
correlation has collapsed to $r(\Delta C, S_T) = +0.09$
[Fig.~\ref{fig:loop}(a),(d)]: the alignment is essentially gone, drifting
weakly negative at later times.  What $\Delta C$ carries at pulse end is
instead information about the \emph{final outcome}, with which it
correlates at $r = +0.80$, the strongest correlation it attains anywhere
along the $t_\mathrm{eval}$ axis, and one that the spurious
forward-ordered construction does not show.  Both correlations then decay: by
$t_\mathrm{eval} = 2$--$3$ each is small and without consistent sign,
as the polynomially growing background OTOC of the Kerr evolution
(Sec.~\ref{sec:intro}; Ref.~\cite{AlmasriReboiro}) progressively buries
the gate's spatial signal.  By $T = 8$ the gate's contribution to
$\Delta C$, while large in magnitude, has lost its preparation-space
structure, its spatial standard deviation is roughly a quarter of its
mean and is uncorrelated with the classical sensitivity.  The bridge
between operator growth and classical sensitivity is therefore real but
transient and \emph{mid-pulse}; the durable pulse-end signal in
$\Delta C$ is its correlation with the outcome, not with the
instantaneous sensitivity.

\paragraph{The Loschmidt echo predicts the final outcome.}
The echo evaluated at $t_\mathrm{eval} = 0.5$ correlates with the final
outcome $P_\mathrm{left}(T = 8)$ at $r = -0.91$,\footnote{The
time-resolved echo correlations of this section are computed on
$\log_{10}F$, the natural scale for a fidelity that ranges over orders
of magnitude.} and at $t_\mathrm{eval} = 1.0$ at $r = -0.94$; the
correlation decays to $-0.51$ by $t_\mathrm{eval} = 3$ as both quantities
saturate.  The sign is the physical content: preparations most disturbed
by the gate (small $F$) are the ones that end most scrambled
($P_\mathrm{left}$ nearest $\tfrac{1}{2}$).  Because the outcome itself
tracks the classical partition
(Sec.~\ref{sec:scrambling}), the echo is partly anticipated by the
classical leak set as well; its value is as a purely quantum,
gate-end early-warning quantity, not as an independent identifier of a
classically invisible set.

\paragraph{The outcome locks in early.}
The correlation between $P_\mathrm{left}(t_\mathrm{eval})$ and
$P_\mathrm{left}(8)$ rises from $0.61$ at $t_\mathrm{eval} = 0.5$ to
$0.87$ at $1.0$ and $0.99$ at $2.0$: within roughly one Kerr time
after the pulse, the final degree of scrambling is determined, and the
remaining evolution only relaxes the state within each well.

\subsection{The diagnostic chain at the reference pulse}
\label{sec:chain}

Combining the above with Secs.~\ref{sec:classical}
and~\ref{sec:quantum}: the classical sensitivity ridge locates the
transport boundary; the gate-induced OTOC carries the same geometric
information only \emph{mid-pulse} ($r = +0.97$ at
$t_\mathrm{eval} = 0.5$) and loses it by pulse end, where it instead
tracks the outcome ($r = +0.80$ at $t_\mathrm{eval} = 1.0$); the
Loschmidt echo in that same pulse-end window predicts the final quantum
outcome ($r = -0.94$); and the outcome itself tracks the classical
partition ($r = +0.61$).  Each link is verified on the same
grid, for the same model and pulse.  How far this picture extends
beyond the reference pulse is the subject of the next section.


\FloatBarrier

\end{document}